\RequirePackage{fix-cm}
\documentclass[twocolumn,epjc3]{svjour3}  
\smartqed  
\RequirePackage{graphicx}
\journalname{Eur. Phys. J. C}

\usepackage{bm}
\usepackage{latexsym}
\usepackage{dcolumn}
\usepackage{amsfonts,amssymb}
\usepackage{graphicx}
\usepackage{epsfig}
\usepackage{psfrag}
\usepackage{nccmath}
\usepackage{moresize}
\usepackage{enumerate}
\usepackage[hang,flushmargin]{footmisc}
\usepackage[titletoc,toc]{appendix}
\usepackage{lipsum}
\usepackage{subcaption}
\usepackage{hyperref}
\usepackage{rotating}
\usepackage{multirow}
\usepackage{mathtools}

\usepackage{cite}
\usepackage{tikz}
\usetikzlibrary{positioning}

\newcommand{\be}{\begin{equation}}
\newcommand{\ee}{\end{equation}}
\newcommand{\bea}{\begin{eqnarray}}
\newcommand{\eea}{\end{eqnarray}}
\newcommand{\bse}{\begin{subequations}}
\newcommand{\ese}{\end{subequations}}
\newcommand{\bce}{\begin{center}}
\newcommand{\ece}{\end{center}}
\newcommand{\bfg}{\begin{figure}}
\newcommand{\efg}{\end{figure}}
\newcommand{\bit}{\begin{itemize}}
\newcommand{\eit}{\end{itemize}}
\newcommand{\bed}{\begin{description}}
\newcommand{\eed}{\end{description}}
\newcommand{\ben}{\begin{enumerate}}
\newcommand{\een}{\end{enumerate}}
\newcommand{\nn}{\nonumber}

\newcommand{\la}{\label}
\newcommand{\pa}{\partial}
\newcommand{\fr}{\frac}
\newcommand{\sq}{\sqrt}
\newcommand{\no}{\noindent}

\def\a  {\alpha}
\def\b  {\beta}
\def\c  {\gamma}
\def\C  {\Gamma}
\def\d  {\delta}

\def\e  {\epsilon}

\def\f  {\phi}

\def\k  {\kappa}
\def\l  {\lambda}
\def\L  {\Lambda}
\def\m  {\mu}
\def\n  {\nu}

\def\O  {\Omega}

\def\r  {\rho}

\def\Th {\Theta}
\def\s  {\sigma}
\def\t  {\tau}
\def\vep {\varepsilon}

\def\z {\zeta}
\def\le {\left}
\def\ri {\right}
\newcommand{\cA}{\mathcal A}

\newcommand{\cL}{\mathcal L}

\newcommand{\cO}{\mathcal O}

\newcommand{\cQ}{\mathcal Q}

\newcommand{\cT}{\mathcal T}

\newcommand{\bcT}{\overline{\mathcal T}}

\newcommand{\nab}{\nabla\!}

\newcommand{\Rt}{\widetilde{R}}

\newcommand{\Ct}{\widetilde{\C}}

\newcommand{\hg}{\widehat{g}}

\newcommand{\hT}{\widehat{T}}
\newcommand{\hX}{\widehat{X}}

\newcommand{\Sm}{S^{(m)}}

\newcommand{\vx}{\vec{\pmb x}}

\newcommand{\Tm}{T^{(m)}}
\newcommand{\Td}{T^{(d)}}

\newcommand{\rmt}{\r^{(m)}}

\newcommand{\rct}{\r^{(c)}}

\newcommand{\rdt}{\r^{(d)}}
\newcommand{\pdt}{p^{(d)}}

\newcommand{\rx}{\r_{\!_X}}
\newcommand{\px}{p_{\!_X}}

\newcommand{\rmp}{\r^{(m)}_{_0}}

\newcommand{\rcp}{\r^{(c)}_{_0}}

\newcommand{\sw}{\mathsf w}

\newcommand{\wx}{\sw_{\!_X}}

\newcommand{\Hp}{H_{_0}}
\newcommand{\tp}{t_{_0}}

\newcommand{\fp}{\f_{_0}}

\newcommand{\betp}{\b_{_0}}
\newcommand{\zetp}{\z_{_0}}

\newcommand{\bp}{\overline{p}}

\newcommand{\bw}{\overline{\sw}}

\newcommand{\brho}{\overline{\rho}}

\newcommand{\bOL}{\overline{\O}^{(\!\L\!)}}
\newcommand{\bOLp}{\overline{\O}^{(\!\L\!)}_{_0}}

\newcommand{\bH}{\overline{H}}

\newcommand*\rfraa[2]{{}^{\displaystyle #1}\!\!\!\diagup_{\!\!\displaystyle #2}}

\newcommand{\bdm}{\begin{displaymath}}
\newcommand{\edm}{\end{displaymath}}

\long\def\symbolfootnote[#1]#2{\begingroup%
\def\thefootnote{\fnsymbol{footnote}}\footnote[#1]{#2}\endgroup}
\numberwithin{equation}{section}
\providecommand{\appendixname}{Appendix}

\newcommand{\diag}{\mbox{diag}}
\newcommand{\Shi}{\mbox{Shi}}

\newcommand{\Yp}{Y_{\!_0}}
\newcommand{\UAp}{U^A_{\!_0}}
\newcommand{\UTp}{U^T_{\!_0}}
\newcommand{\eYp}{\vep^Y_{\!_0}}
\newcommand{\eAp}{\vep^A_{\!_0}}
\newcommand{\eTp}{\vep^T_{\!_0}}

\newcommand{\bW}{\overline{W}}
\newcommand{\sgp}{\s_{\!_0}}
\newcommand{\mup}{\m_{_0}}
\newcommand{\Smmt}{S_{_{MMT}}}
\newcommand{\Rin}{\!~^{(3)}\!R}

\begin{document}


%
%

\title{Mimetic-Metric-Torsion with induced Axial mode and Phantom barrier crossing}
%
\author{Sourav Sur\thanksref{e1} \and Ashim Dutta\thanksref{e2} \and Hiyang Ramo 
Chothe\thanksref{e3}
}                     
\thankstext{e1}{sourav.sur@gmail.com}
\thankstext{e2}{ashim1921@gmail.com}
\thankstext{e3}{rmo.tnm@gmail.com}
\institute{Department of Physics and Astrophysics \\
University of Delhi, New Delhi - 110 007, India}
\date{Received: date / Revised version: date}
\maketitle

\begin{abstract}
We extend the basic formalism of mimetic-metric-torsion gravity theory, in a way
that the mimetic scalar field can manifest itself geometrically as the source of
not just the trace mode of torsion, but its axial (or, pseudo-trace) mode as well.
Specifically, we consider the mimetic field to be (i) coupled explicitly to the
well-known Holst extension of the Riemann-Cartan action, and (ii) identified with 
the square of the associated Barbero-Immirzi field, presumably a pseudo-scalar. 
The conformal symmetry originally prevaling in the theory would still hold, as 
the associated Cartan transformations do not affect the torsion pseudo-trace, 
and hence the Holst term. Demanding the theory to preserve the spatial parity 
symmetry as well, we show a geometric unification of the cosmological dark sector, 
and the feasibility of a super-accelerating regime in the course of evolution of 
the universe. From the observational perspective, assuming the cosmological 
evolution profile to be very close to that for $\L$CDM, we further illustrate a 
smooth crossing of the so-called phantom barrier at a low red-shift, albeit 
with a restricted parametric domain. Subsequently, we determine the extent of 
the super-acceleration by examining the evolution of the relevant torsion 
parameters. 
\keywords{Torsion \and Dark Energy theory \and Modified Gravity \and Cosmology 
of theories beyond the SM}
\end{abstract}
%



\section{Introduction}

Emergence of the theory of mimetic gravity
\cite{CM-mm,CMV-mm,SVM-mm}
has marked a fascinating recent advancement in modified gravity researches
purporting to a geometric description of the cosmic dark sector constituents, 
viz. {\em dark energy} (DE) and {\em dark matter} (DM) 
\cite{CST-rev,AT-book,wols-ed,MCGM-ed,BCNO-rev,chiba-mg,NO-mg1,NO-mg2,NO-mg3,
FTT-mg,SF-mg,FT-mg,clift-mg,papa-ed,NOO-rev}.
Such a theory has attracted a lot of attention not only for providing an exact 
mimicry of a dust-like cold dark matter (CDM) component in the standard framework
of Friedmann-Robertson-Walker (FRW) cosmology, but also for arousing, via simple 
extensions, the very prospect of the unification of the entire dark sector. 
In fact, the original Chemseddine-Mukhanov (CM) model
\cite{CM-mm}
of mimetic gravity\footnote{See also some relevant precursors
\cite{LSV-dde,GGWC-dde,CMNO-dde}.}
have had a plethora of extensions proposed and studied from various perspectives 
till date
\cite{CMV-mm,SVM-mm,CP-mhd,GMF-mhd,CKOT-minst,IRS-minst,HNK-minst,ZSML-minst,TK-minst,
FGM-minst,BLN-mdhost,LMNV-mdhost,MS-mnfR,MMG-mfRv,AOO-mfRv,BHHA-mfRv,ABKM-mh,CMSVZ-mh,
BCC-mbi,CBC-mbi,SN-mbw,YYZL-mbw,ZZGL-mbw,CM-mmg1,CM-mmg2,MS-mmg,SVA-mmg,MAM-mfR,
MSVZ-mhor,kosh-mhd,ST-m2s,KNY-mvt,GMF-mgf,GMFM-mgf,JV-mgf,CMR-maf,CMR-mhor,MS-mhor},
with many interesting implications in both cosmology and astrophysics
\cite{barv-mm,HV-mm,RMMM-mdyn,DKSTS-mdyn,LS-mdyn,OO-mdyn,NO-mfR,NOO-mfR,MSV-mfR,
OO-mfR1,OO-mfR2,CM-msing,BGY-msing,HSP-msing,MSVZ-mgal,vag-mgal,MS-mco,MMGM-mco,
AO-mco,CM-mco,nas-mco,NHB-mco,SJ-mgw,BBFLMS-mgw,LSYN-mgw,BPT-mgw,RSCV-mgw,
GBKM-mgw,CRSV-mgw,CMR-bh,HJV-axioncc}.

Strictly speaking, the basic CM theory amounts to a scalar-tensor reformulation 
of General Relativity (GR), of a specific sort which exploits the diffeomorphism
invariance to reparametrize the physical metric $g_{\m\n}$ by a fiducial metric 
$\hg_{\m\n}$ and a scalar field $\f$, in a way that $g_{\m\n}$ remains invariant 
under a conformal transformation of $\hg_{\m\n}$. The gravitational conformal 
degree of freedom thus gets encoded by the field $\f$, known as the `mimetic 
field'
\cite{CM-mm},
which ostensibly has no prior relevance to geometry however. It therefore seems
legitimate to look for a generalized mimetic scenario in which $\f$ can manifest 
itself geometrically, say for instance, as the source of {\em torsion}, which is 
often regarded as important a space-time characteristic as curvature\footnote{Refer
to the hefty literature on the long history torsion gravity theories (from the 
pioneering Einstein-Cartan formulation to the plethora of scenarios that emerged 
from supergravity, string theory, brane-worlds, etc., or the modified 
{\it tele-parallelism}, Poincar\'e gauge gravity, and so on)
\cite{einst-rel,traut,HVKN-trev,akr-tbook,SG-tbook,SS-tbook,HDMN-trev,HO-trev,
shap-trev,blag-book,fab-thesis,CL-extgrav,popl-trev,WZ-trev,mus-MTconf,fab-MTconf,
berg-MTconf,pmssg,ham,bmssssg-kr1,BC-kr,bmssssg-kr2,FF-fT,BF-fT,LSB-fT,CCDDS-fT,
BMT-fT,CCLS-fT,BB-fT,BCFN-fT,YN-PGT,MGK-PGT,NSV-PGT,BHN-PGT,GLT-PGT,HB-PGT,LC-PGT,
obu-PGT,HRR-proptor,CF-proptor,saa-proptor,BS-proptor,popl-proptor,BC-proptor,
ND-proptor,ACCF-extgr,FC-extgr,CMT-extgr,HORB-skew,FRM-scaltor,VFS-sqtor,lu-sqtor,
fab-sqtor,VCVM-sqtor,KS-deg1,KS-deg2,SS-deg1,KS-deg3,SS-deg2,ssasb-mst1,ssasb-mst2,
ssasb-mst3,MO-mfT,GZYSL-mfT,IMMSV-mfT}. 
See also the literature on the predicted observable effects of torsion, and their 
extensive searches
\cite{ROH-skew,Ni-skew,ssgss-kr1,skpmssgas-kr,skpmssgss-kr,
skssgss-kr,ssgss-kr2,dmssgss-kr1,sssdssg-kr,AMT-kr,CMS-kr,ssgss-kr3,dmssg-kr,
dmssgss-kr2,sdadssg-kr,adbmssg-kr,scssg-kr1,scssg-kr2,scssg-kr3,GLSW-fT,GLS-fT,
IS-fT,CGSV-fT,SST-fT,KPS-fT,CLPR-fT,FSGS-fT,SNY-PGT,mink-PGT,NRR-PGT,KRT-texpt,
FR-texpt,BF-texpt,HOP-texpt,CCR-texpt,CCSZ-texpt,LP-texpt}.}.
Such a generalization essentially implies contemplating on a conceivable way of 
expanding the purview mimetic gravity to the metric-compatible Riemann-Cartan 
($U_4$) geometry that admits torsion in addition to curvature. An endeavour 
pertaining to this has resulted in our mimetic-metric-torsion (MMT) gravity 
formulation in a recent paper (henceforth, `Paper\,1'
\cite{RAS-MMT}),
with the eventual enunciation of a viable unified cosmological dark sector paradigm. 
Nevertheless, scope remains for looking into the subtler aspects of the evolving dark 
sector, by suitably extending the MMT theory, which we intend to do in this work. 

From the technical point of view, a consistent MMT formulation primarily amounts to a 
proper ratification of the principle of mimetic gravity in presence of torsion. This 
specifically implies the isolation of the conformal (scalar) degree of freedom of 
gravity via not only the parametrization of the physical metric $g_{\m\n}$ by a 
fiducial metric $\hg_{\m\n}$ and the mimetic field $\f$, but also that of the physical 
torsion $T^\a_{~\m\n}$ by a corresponding fiducial space torsion $\hT^\a_{~\m\n}$ and 
$\f$. We have been able to assert this torsion parametrization in Paper\,1 by 
examining carefully the form of the fiducial space Cartan transformations under which,
and a conformal transformation of $\hg_{\m\n}$, the physical fields $g_{\m\n}$ and 
$T^\a_{~\m\n}$ are to be preserved. Considering next, from certain standpoints, a 
contact coupling $\b(\f)$ of the mimetic field and the torsional part of the $U_4$ 
Lagrangian, it has been shown that the trace mode $\cT_\m$ of torsion gets sourced by 
$\f$, which thereby manifests itself geometrically
\cite{RAS-MMT}.
The resulting energy-momentum tensor resembles that of a perfect fluid\footnote{Or an 
imperfect fluid, if one also considers incorporating term(s) containing higher order 
derivatives of $\f$, such as $(\square \f)^2$, in the effective Lagrangian
\cite{HV-mm,SVM-mm,MV-mm,CR-mm,ramz-mm,BR-mm}.
}, dubbed the `MMT fluid', which characterizes {\em dust}, albeit with a non-zero 
pressure. The latter is attained by virtue of an effective potential $W (\f)$, 
culminating from $\b (\f)$ and its first derivative, in an equivalent formulation 
illustrated in Paper\,1. Any desired expansion history of the universe can therefore 
be reconstructed in the standard FRW cosmological framework, by suitably choosing the 
form of the `MMT coupling' function $\b (\f)$. However, in most circumstances such 
choices lack proper physical motivation, and are merely phenomenological. In Paper\,1 
though, the consideration of a well-known (and well-motivated) coupling $\b (\f) \sim 
\f^2$ has made the potential $W$ identifiable as a cosmological constant $\L$ (modulo 
a dimensional factor). The ultimate outcome have therefore been an effective $\L$CDM 
evolution in the MMT cosmological setup, even when the torsion strength diminishes to 
its unobtrusiveness at late times (thus corroborating to its miniscule experimental 
signature till date
\cite{KRT-texpt,FR-texpt,BF-texpt,HOP-texpt,CCR-texpt,CCSZ-texpt,LP-texpt}).

Nevertheless, the entire MMT formalism in Paper\,1 is in some sense {\it minimalistic}, 
since it does not reflect on the actual (intriguing) characteristics of torsion in 
shaping up the cosmological evolution profile. Specifically, the latter has no direct 
influence of the antisymmetry property of torsion, unless one assigns an external 
source for the {\em axial} (or {\em pseudo-trace}) mode $\cA_\m$ of torsion. Such an 
external source, supposedly a non-gravitational field, is anyhow not much interesting, 
since its admittance may not only affect the predictability of the theory (due to the 
additional degree(s) of freedom), but also imply the redundancy in aspiring for a 
geometric unification of the cosmological dark sector. It is imperative therefore to 
look for extending the basic MMT formalism in such away that apart from torsion's trace 
$\cT_\m$, its axial mode $\cA_\m$ (with no assigned external source) can actively 
influence the dynamical solutions of equations of motion in a given setup (that of FRW 
cosmology for instance). A simple and natural way to do so is to consider the mimetic 
field's explicit coupling with not only the torsional part of the $U_4$ curvature 
scalar $\Rt$, but also its Hodge dual. The latter is commonly known as the `Holst 
extension' of the $U_4$ Lagrangian, describing modified versions of the conventional 
metric-torsion theories of Einstein-Cartan (EC) type, or more generally, the refined 
editions of the canonical (Hilbert-Palatini) connection-dynamic gravitational theories
\cite{HMS-pvtor,holst-pvtor,imm-Holst,rov-Holst,PR-Holst,thiem-Holst,kaul-Holst,
ban-Holst,PS-Holst,ST-Holst,RV-Holst}.
In fact, it had been a generalized Hilbert-Palatini formulation of gravity which had 
accounted for the Holst term in Holst's original paper in 1996
\cite{holst-pvtor}.
Nevertheless, the equivalent term, viz. the Hodge dual of $\Rt$, in the 
Einstein-Hilbert formulation had long been known for
\cite{HMS-pvtor}.
In either formulation though, the Holst extension being merely {\em topological} (at 
least, at the minimal level), does not affect the classical dynamics. However, apart 
from a potential significance in canonical quantum gravity (particularly, relating to 
Ashtekar-Barbero formulation)
\cite{imm-Holst,rov-Holst,PR-Holst,RV-Holst,thiem-Holst,asht1-QGR,asht2-QGR,barb-QGR,
AL-QGR,mer-QGR},
the Holst term's presence may reveal intriguingly via {\em gravitational parity 
violation}, possible consequences of which have been studied in various contexts, such 
as that of an axial torsion induced by the string theoretic Kalb-Ramond field
\cite{bmssgss-pvtor,bmssssgss-pvtor,dmpmssg-pvtor,CM-pvtor,FMT-pvtor,mer-pvtor}. 
 
A majority of recent works on the formulations of metric-torsion or connection-dynamic 
theories have pondered on relaxing the topological characterization of the Holst 
extension, in which case it can in principle affect the classical dynamics. An
appropriate methodology, which has been motivated from various standpoints (such as 
the chiral anomaly cancellation), is to promote the associated coupling parameter $\c$, 
known as the Berbero-Immirzi (BI) parameter, to the status of a scalar or a 
pseudo-scalar field
\cite{mer-QGR,mer-pvtor,CM-BI,MT-BI,BCM-BI,BM-BI,BM1-BI}.
The latter of course is the most suitable option if the theory is to preserve spatial 
parity symmetry --- a consideration amenable to our endeavour on extending the basic
MMT formalism by incorporating the Holst term, coupled to the mimetic field $\f$, in 
this paper. In fact, a pseudo-scalar (or, an {\em axion}-like) BI field $\c$ is 
reasonable from the point of view of ensuring not only that the Holst term would 
affect the MMT cosmological solutions irrespective of the $\f$-coupling, but also 
that such solutions would conform to the general non-perception of the effect(s) of 
gravitational parity violation, at least at the background level\footnote{Note for 
e.g. that the recent Planck observations do not even provide a reasonably clear
indication of the $E$-$B$ and $T$-$B$ cross-correlations in the Cosmic Microwave 
Background (CMB) polarization power spectra
\cite{Planck-PV}.}.
It is also worth mentioning here that the admittance of the Holst term does not 
disturb the overall conformal symmetry of the MMT theory, subject to the torsion 
parametrization suggested in Paper\,1, as the associated Cartan transformations 
leave the axial mode $\cA_\m$ of torsion unaffected
\cite{RAS-MMT}.

Following are the additional steps we take in our extended MMT formulation:
\ben
\item  Using a Lagrange multiplier, we make an identification of $\f$ with $\c^2$, 
modulo some constant. This of course is meant for a simplification, not only from 
the point of view of dealing effectively with just a solitary scalar field $\f$, 
but also in letting $\f$ manifest itself geometrically as the source of both the 
modes $\cT_\m$ and $\cA_\m$ of torsion. Moreover, note that for such simplification, 
alongside the preservation the gravitational parity symmetry, 
it suffices to identify $\f$ with any even function of the pseudo-scalar BI field 
$\c$, the $\c^2$ being just the simplest possible choice.
\item We retain the overall MMT coupling $\b (\f)$, but now between $\f$ and the 
entire torsional part of the extended $U_4$ Lagrangian (the Holst term inclusive). 
This is essential in the sense that we require to get an equation of motion 
leading to $\cT_\m \propto \pa_\m \f$ (as in Paper\,1), so that the MMT fluid 
continues to be dust-like\footnote{That is, the fluid velocity $u_\m$ (which is 
taken to be equal to $\pa_\m \f$, in analogy with k-essence cosmologies
\cite{AMS2000-kess,AMS2001-kess,MCLT-kess,schr-kess,sssd-kess})
has to be tangential to the time-like affine geodesics (or the {\em auto-parallels}) 
in the $U_4$ space-time. More specifically, the fluid acceleration must vanish, and 
thereby confirm the auto-parallel equation (see the Appendix of Paper\,1
\cite{RAS-MMT}).}
even in the extended setup. Furthermore, we consider the same form of the coupling
function, viz. $\b (\f) \sim \f^2$, as in Paper\,1, for the motivations cited 
therein. In fact, such a quadratic coupling can have a natural appearance in MMT
gravity, as demonstrated in our subsequent paper
\cite{MMT-DS}.
\item We also consider augmenting the effective MMT action with a higher derivative 
$(\square \f)^2$ term, which leads to a non-zero sound speed $c_s$ of mimetic matter 
density perturbations, side-by-side preserving the form of the background cosmological 
solution one obtains in its absence
\cite{CMV-mm,SVM-mm}.
\een
With this setup, we derive the MMT field equations and constraints, working out in 
due course the form of the effective potential $W (\f)$ in section\,\ref{sec:genPMT}.
 
Subsequently, in section\,\ref{sec:cosmPMT} we carry out the study of the qualitative 
aspects of the scenario that emerges in the realm of the standard FRW cosmology. 
In particular, we show that equation of state parameter $\wx$ of the effective DE 
component, construed in such a scenario, always tends to fall below the $\L$CDM value 
$-1$, thereby exhibiting a {\em super-accelerating} phase of cosmic expansion. The 
transition from $\wx > -1$ to $\wx < -1$, or the so-called {\em phantom barrier} 
crossing, is nonetheless an intriguing pathology, with a fair amount of observational 
support. For instance, the results of the Planck 2018 combined analysis
\cite{Planck18}
do conform to a certain extent the pre-existing signature of such a crossing at an 
epoch $t_{_C}$ in the recent past phase of evolution of the universe, and that the  
super-accelerated expansion is still continuing (albeit mildly) at the present 
epoch $\tp$. However, commonly known scalar field DE models (quintessence, k-essence, 
etc.) cannot usually account for this without the involvement of ghost (or phantom) 
degree(s) of freedom. There is no such concern though, with our MMT extension scheme 
outlined above, as the corresponding Lagrangian we propose does not ostensibly consist 
of any ghost-like term, and in a reduced form looks precisely the same as that in the 
existing literature on mimetic gravity
\cite{CMV-mm,SVM-mm}.
It is the specific form of the potential $W (\f)$ we get as a consequence of our weird 
constraining of the system, that leads to the phantom crossing. Actually, the total 
(mimetic + external) energy-momentum content of the system does not violate any of the 
energy conditions, except the strong one. It is the mathematical artifact of $W (\f)$ we 
interpret as the effective DE constituent in analogy with conventional multi-component 
cosmological models, that behaves as a phantom. Such an interpretation, although not 
meant for pinpointing any physical DE candidate, helps us immensely in a bit by bit 
understanding of the emergent cosmological scenario. 

In section\,\ref{sec:PMTphantcross}, we demonstrate the physical realizability of the 
phantom crossing, i.e. its occurence at an epoch $t_{_C}$ close to but earlier than 
$\tp$. However, this requires our MMT model parametric values to be kept within very 
narrow domains, which we illustrate after explicitly working out $\wx (t)$, albeit for 
an approximation justifiable from the point of view of the observational concordance 
on $\L$CDM. Specifically, we presume the $\f$-coupled Holst term is not capable of 
producing much distortion of the $\L$CDM solution we have had in its absence in 
Paper\,1. We determine the validity of the approximation (and subsequently the extent 
of the super-acceleration) in section\,\ref{sec:PMT-TA}, by examining the evolution 
profiles of the torsion parameters, viz. the norms of $\cT_\m$ and $\cA_\m$, for 
specific parametric choices. 

We conclude in section \ref{sec:concl} with a summary and a discussion on some relevant 
areas open for future investigations, and thereafter demonstrate in the Appendix, 
the stringent problem of the Ostrogradsky instabilties due to the higher derivative(s), 
and the possible avoidance of the same, via an alternative Lagrangian extension.

As to the notation and conventions, we adopt those in Paper\,1, except using for 
brevity the units in which the gravitational coupling factor $\k^2 \equiv 8 \pi G_N = 1$.

\section{General formalism} \label{sec:genPMT}

To begin with, let us refer the reader to the main tenets of Paper\,1 --- in 
particular, the sections $2$ and $3$ therein, starting with the definition of
the torsion tensor 
\be \label{tor}
T^\a_{\,\, \m\n} :=\, \Ct^\a_{\,\, \m\n} - \Ct^\a_{\,\, \n\m} \,\,,
\ee
in four-dimensional Riemann-Cartan ($U_4$) space-time, characterized by an asymmetric 
(but metric compatible) affine connection $\Ct^\a_{\,\, \m\n}$. We look to formulate 
an extended MMT theory in which the (dimensionless) mimetic scalar field $\f$ can 
drive the classical dynamics, while inducing both the vector modes of torsion, viz. 
its trace $\cT_\m := T^\n_{\,\, \m\n}$ and pseudo-trace $\cA^\s := \e^{\a\b\c\s} 
T_{\a\b\c}$. For this purpose we resort to the following minimal level generalization 
of the basic MMT action (of Paper\,1): 
\bea \label{PMT-ac}
S &=& \Sm +\, \fr 1 2 \int \! d^4 x \sq{- g} \bigg[R (g_{\m\n}) \,+\, 
\a \le(\square \f\ri)^2 \nn\\
&& \quad+\, \b(\f) \bigg\{\Th (g_{\m\n}, T^\a_{~\m\n}) + 
\fr 1 {2 \c} \, {^*\Th}(g_{\m\n}, T^\a_{~\m\n})\bigg\} \nn\\
&& \quad+\, \l \cdot (X - 1) +\, \n \cdot \!\le(\f - s \c^2\ri) \bigg] \,,
\eea
where $\Th$ and ${^*\Th}$ denote the torsion-dependent part of the $U_4$ curvature 
scalar $\Rt$ and the Hodge dual of the $U_4$ curvature tensor $\Rt_{\a\b\c\d}$,
respectively. In explicit forms, they are expressed in terms of $\cT_\m$, $\cA_\m$, 
and the (pseudo)tracefree mode $\cQ^\a_{\,\, \m\n}$ of torsion, as
\bea 
\Th (g_{\m\n}, T^\a_{~\m\n}) &:=& \Rt (g_{\m\n}, T^\a_{~\m\n}) \,-\, R (g_{\m\n}) \nn\\
&=& - 2 \nab_\m \cT^\m -  \mfrac 2 3 \cT_\m \cT^\m + \mfrac 1 {24} \cA_\m \cA^\m \nn\\
&& +\, \mfrac 1 2 \cQ_{\a\m\n} \cQ^{\a\m\n} \,, \label{T-Lagr} \\
^\star \Th (g_{\m\n}, T^\a_{~\m\n}) &:=& \e^{\a\b\c\d} \Rt_{\a\b\c\d} (g_{\m\n}, 
T^\a_{~\m\n}) \nn\\
&=& - \nab_\m \cA^\m - \mfrac 2 3 \cT_\m \cA^\m \nn\\
&& +\, \mfrac 1 2 \e^{\a\b\c\d} \cQ^\m_{~\a\b} \cQ_{\m\c\d} \,, \label{PsT-Lagr}
\eea
with $R$ being the usual Riemannian ($R_4$) curvature scalar, and the tensor indices 
being raised or lowered using the physical metric $g_{\m\n}$. 

The other constituent terms and symbols in Eq.\,(\ref{PMT-ac}) are as explained as
follows: 
%
\bit
\item $\Sm$ is the external matter action.  
\item $\, \square \equiv g^{\m\n} \nabla_\m \nabla_\n$, with $\nabla_\m$ denoting the 
$R_4$ covariant derivative. 
\item $\a$ is a dimensionless constant parameter measuring the strength of coupling of 
$\le(\square \f\ri)^2$. 
\item $\b(\f)$ denotes the overall contact coupling of $\f$ with $\Th \,$, and its 
{\em Holst} extension, viz. $\mfrac {^*\Th} {2\c}  \,$, with $\c$ being the 
(dimensionless) Berbero-Immirzi (BI) field, which is taken to be a pseudo-scalar 
(or an {\it axion}) in order that the intrinsic parity in the theory is 
preserved\footnote{In particular, following the convention of Rovelli {\it et.\,al.}
\cite{rov-Holst,PR-Holst,RV-Holst},
we keep $\c$ in the denominator of the coefficient of $\,^\star \Th$. The factor of 
$2$, also appearing in the denominator, is just for the purpose of the eliminating 
certain cumbersome numerical factors in our subsequent derivations.}. 
\item  $\l$ and $\n$ are scalar Lagrange multiplier fields, used respectively for 
enforcing the {\em mimetic constraint}\footnote{Or, the condition one gets by 
demanding the physical metric $g_{\m\n}$ to be non-singular, and hence invertible.}:
\be \label{mm-constraint}
X \equiv\, -\, g^{\m\n} \, \pa_\m \f \, \pa_\n \f =\, 1 \,\,,
\ee
and for identifying $\f$ with the squared BI field, i.e.
\be \label{PMT-id}
\f \equiv \, s \, \c^2 \,\,, 
\ee
where $s$ is a dimensionless coupling constant.
\eit
Note also the following:
\ben
\item The above action (\ref{PMT-ac}) has merits in itself from the point of view that 
it comprises of curvature and torsion constructs of only the lowest orders (linear and 
quadratic respectively). So there is apparently no scale dependence in the theory.
\item The overall conformal symmetry of the theory could be retained even in presence 
of the $\f$-coupled Holst term, viz. $\mfrac{\b (\f)}{2 \c} \,^\star\Th \, \sq{-g}\, $, 
in the Lagrangian. In principle, just as in Paper\,1, we may resort to the physical 
metric and torsion parametrizations:
\bea 
&& g_{\m\n} =\, \hX \, \hg_{\m\n} \,\,, \label{MMT-param-g} \\
&& T^\a_{~\m\n} =\, \hT^\a_{~\m\n} +\, q \,\d^\a_{[\m} \, \pa_{\n]} (\ln \hX) \,\,,
\label{MMT-param-T}
\eea 
where $\, \hX = - \,\hg^{\m\n} \, \pa_\m \f \, \pa_\n \f \,$ and $q$ is a real 
numerical parameter, which although arbitrary in general, had been set to unity from 
certain standpoints in Paper\,1 (see the section $2$ therein
\cite{RAS-MMT}).
Under the conformal and Cartan transformations of the fiducial metric and torsion, viz.
\bea 
&& \hg_{\m\n} \rightarrow\, e^{2\s} \, \hg_{\m\n} \,\,, \label{conf-aux-g} \\
&& \hT^\a_{~\m\n} \rightarrow\, \hT^\a_{~\m\n} +\, q \le(\d^\a_\m \, \pa_\n \s - 
\d^\a_\n \, \pa_\m \s\ri)\,, \label{conf-aux-T}
\eea
$g_{\m\n}$ and $T^\a_{~\m\n}$ remain invariant, for any given real scalar function 
of coordinates $\s$. Now, the conformal covariance of the physical torsion (i.e. its 
conformal invariance in the mixed form $T^\a_{~\m\n}$) implies the same for its 
irreducible modes as well. In particular, the transformations (\ref{conf-aux-g}) and
(\ref{conf-aux-T}) leave invariant such modes in the respective forms $\, \cT_\m$, 
$\cA_\m$ and $\cQ^\a_{~\m\n}$. Moreover, a given conformal transformation would not 
affect the scalar field $\f$, as it is dimensionless\footnote{Otherwise, it would be 
$\f \rightarrow e^{-\s} \f$, if $\f$ happens to be a unit mass dimension scalar 
field, as in most theories
\cite{mal-CG}.}.
By the same token, a conformal transformation would leave the pseudo-scalar $\c$ 
invariant, since $\c$ is dimensionless as well. Therefore, expressing the 
$\f$-coupled Holst term explicitly as
\bea \label{phi-Holst}
\fr{\b (\f)}{2 \c} \,^\star\Th \, \sq{-g} &=& - \, \fr{\b (\f)}{2 \c} \bigg[\pa_\m 
\le(\sq{-g} \, g^{\m\n} \cA_\n\ri) \nn\\
&& ~+ \sq{-g} \bigg\{\mfrac 2 3 \, g^{\m\n} \cT_\m \cA_\n \nn\\
&& ~- \mfrac 1 2 \, \e^{\a\b\c\d} g_{\m\n} \cQ^\m_{~\a\b} \cQ^\n_{~\c\d} 
\bigg\}\bigg],
\eea
we can see its invariance under (\ref{conf-aux-g}) and (\ref{conf-aux-T}), since all 
of its elements, including the contravariant Levi-Civita tensor density $\vep^{\a\b\c\d} 
\equiv \sq{-g}\, \e^{\a\b\c\d}$, remain unchanged.
\item The explicit $\f$-coupling with the Holst term may not seem to be necessary for 
the MMT extension we are seeking. In fact, it may seem much convenient to consider the 
action
\bea \label{PMT-ac-alt}
S =\, \Sm &+& \fr 1 2 \int \! d^4 x \sq{- g} \bigg[R +\, \a \le(\square \f\ri)^2 \nn\\
&& \,+\, \b(\f) \, \Th \,+\, \fr 1 {2 \c} \, {^*\Th} \nn\\
&& \,+ \l \, (X - 1) + \n \le\{\f - f(\c)\ri\} \bigg] \,.
\eea
This, unlike (\ref{PMT-ac}), does not contain any such coupling, and invokes, instead
of (\ref{PMT-id}), the constraint $\f = f (\c)$, where $f (\c)$ needs to be an even 
function of the axionic BI field $\c$, in order to preserve intrinsic parity. We 
however, prefer to stick to the action\,(\ref{PMT-ac}), not only for some technical 
simplifications, but also for clarity in understanding certain results. Consider for 
instance the torsion mode $\cA_\m$, which is the most crucial ingredient in this 
extended MMT setup. As shown below (see Eqs.\,(\ref{A-constr})), the 
action\,(\ref{PMT-ac}) leads to the result that $\cA_\m$ is determined completely 
by the dynamical solution of the BI field $\c$, regardless of the chosen form of the 
coupling function $\b (\f)$. On the other hand, it can be easily verified that 
$\cA_\m$ derived from the action\,(\ref{PMT-ac-alt}) would depend not only on $\c$ 
(and $\pa_\m \c$), but also on $\b (\f)$. With either of the actions though, the 
determination of $\cA_\m$ would ultimately require the solution for $\f$, once the 
latter is pre-assigned to be identified with some even function of $\c$ (for e.g. 
$\c^2$). Nevertheless, the clarity of wherefrom a given observable (such as the 
norm of $\cA_\m$) emerges is somewhat obscured if one uses (\ref{PMT-ac-alt}). 
\een
Eliminating surface terms, we have the action (\ref{PMT-ac}) expressed in the form
\bea \label{PMT-ac1}
S &=& \Sm \,+\, \fr 1 2 \int \!\! d^4 x \sq{- g} \bigg[R (g_{\m\n}) \nn\\
&&\quad+\, \a \le(\square \f\ri)^2 + \l \, (X - 1) + \n \le(\f - s \c^2\ri) \nn\\
&&\quad+\, \b(\f) \bigg\{2 \le(\ln \b\ri)_\f \cT_\m \pa^\m \f 
-\, \mfrac 2 3 \cT_\m \cT^\m \nn\\
&&\qquad\qquad+\, \mfrac 1 {24} \cA_\m \cA^\m 
+\, \mfrac 1 2 \cQ_{\a\m\n} \cQ^{\a\m\n} \bigg\}\nn\\
&&\quad+\, \fr{\b(\f)}{2 \c} \bigg\{\le[\le(\ln \b\ri)_\f \pa_\m \f -\, \pa_\m 
\le(\ln \c\ri)\ri] \cA^\m \nn\\
&&\qquad\qquad-\, \mfrac 2 3 \cT_\m \cA^\m +\, \mfrac 1 2 \e^{\a\b\c\d} 
\cQ^\m_{~\a\b} \cQ_{\m\c\d} \bigg\}\bigg] \,,
\eea
where $\, \le(\ln \b\ri)_\f \equiv \mfrac d {d\f} \le(\ln \b\ri)$, and the indices 
have been raised and lowered using the physical metric $g_{\m\n}$. 

Assuming that no external sources of the torsion modes $\cT_\m, \cA_\m$ and 
$\cQ^\a_{\,\, \m\n}$ exist in the matter action $\Sm$, the variation of 
(\ref{PMT-ac1}) with respect to the each of them leads us to the constraints 
\bea  
&& \cT_\m =\, \fr 3 2 \le[\le(\ln \b\ri)_\f \pa_\m \f -\, \fr{\pa_\m \c}{\c \le(\c^2 
+ 1\ri)}\ri] \,, \label{T-constr} \\
&& \cA_\m =\, \fr {6 \, \pa_\m \c}{\c^2 + 1} \,\,, \label{A-constr} \\
&& \cQ^\a_{\,\, \m\n} = 0 \quad \le(\forall \,\, \a,\m,\n\ri) \,\,, \label{Q-constr}
\eea
showing that $\cA_\m$ is fully dependent on the dynamics of the BI field $\c$, 
whereas $\cT_\m$ is not so. Ultimately of course, the identification $s \c^2 = 
\f \,$ [{\it cf}. Eq.\,(\ref{PMT-id})] would leave both these modes, $\cT_\m$ 
and $\cA_\m$, as functions of $\f$ and $\pa_\m \f$ only. In fact, we may justify 
such an identification from the point of view of what it entails, viz. the 
condition $\, \cT_\m \propto \pa_\m \f \,$ necessary for the MMT fluid to remain 
dust-like even in presence of torsion (as proved explicitly in the Appendix of 
Paper\,1
\cite{RAS-MMT}).
Accordingly, the norms of $\cT_\m$ and $\cA_\m$ can finally be expressed as
\bea 
&&\!\!\! \cT^2 \equiv - g^{\m\n} \cT_\m \cT_\n = \fr{9 X}{4 \f^2} \bigg[\f 
\le(\ln \b\ri)_\f - \fr s {2 (\f + s))}\bigg]^2 , \label{T-evol} \\ 
&&\!\!\! \cA^2 \equiv\, - g^{\m\n} \cA_\m \cA_\n = \fr{9 s X}{\f (\f + s)^2} \,,
\label{A-evol} 
\eea
where $\, X \equiv - g^{\m\n} \pa_\m \f \, \pa_\n \f$. 

Furthermore, choosing the same quadratic coupling as in Paper\,1 (for the motivations 
cited therein), viz.
\be \label{PMT-coup}
\b (\f) \,=\, \betp \le(\fr{\f}{\fp}\ri)^2 \,\,, 
\ee
with $\, \fp, \betp$ as some constant reference values, we may reduce the action 
(\ref{PMT-ac}) to 
\bea \label{PMT-ac2}
S =\, \Sm &+& \fr 1 2 \int \! d^4 x \sq{- g} \bigg[R +\, \a \le(\square \f\ri)^2 \nn\\
&&\qquad \qquad +\, \l \, (X - 1) -\, W(\f) \, X \bigg] \,.
\eea
This is of the same form as in Paper\,1, albeit with the effective potential 
\be \label{PMT-W}
W(\f) =\,  2 \L \Big[1 -\, Y (\f) \Big] \,, 
\ee
where
\be \label{PMT-YL} 
Y (\f) =\, \fr s {16 \le(\f + s\ri)} \qquad \mbox{and} \qquad 
\L =\, \fr{3 \betp}{\fp} \,.
\ee
Because of the mimetic constraint $X = 1$ [{\it cf}. Eq.\,(\ref{mm-constraint})], 
the gravitational field equations are therefore,
\be \label{PMT-eq1}
R_{\m\n} - \mfrac 1 2 g_{\m\n} R \,=\, \k^2 \le[\Tm_{\m\n} \,+\, \Td_{\m\n}\ri] \,,
\ee
where $R_{\m\n}$ is the usual ($R_4$) Ricci tensor, $\Tm_{\m\n}$ is the matter 
energy-momentum tensor and $\Td_{\m\n}$ is its mimetic extension. The latter is 
given by
\bea \label{em-d}
\Td_{\m\n} &=& \Big[R + \Tm - 2 W\Big] \pa_\m \f \, \pa_\n \f 
\,- \mfrac 1 2 g_{\m\n} W \nn\\
&&\, +\, 2\, \a \le(J_{\m\n} +\, J \, \pa_\m \f \, \pa_\n \f\ri) \,,
\eea
where $\Tm$ denotes the trace of $\Tm_{\m\n}$, and
\bea 
J_{\m\n} &:=& \mfrac 1 2 \Big[\pa_\m (\square \f) \, \pa_\n \f +\, \pa_\n 
(\square{\f}) \, \pa_\m \f\Big] \nn\\
&&-\, \mfrac 1 2 g_{\m\n} \Big[\pa_\s (\square \f) \, 
\pa^\s \f + \mfrac 1 2 (\square \f)^2\Big] \,, 
\label{J-def}\\
J &=& g^{\m\n} J_{\m\n} =\, - \Big[\pa_\s (\square \f) \, \pa^\s \f + (\square 
\f)^2\Big] \,. 
\label{J-tr}
\eea
On the other hand, the variation of the action (\ref{PMT-ac2}) with respect to $\f$ 
leads to the field equation:
\bea \label{PMT-eq2}
&& \nabla_\m \le[\le(R + \Tm - 2 W\ri) \pa^\m \f\ri] \nn\\
&&\qquad =\, \mfrac 1 2 W_\f -\, \a \Big[\square (\square \f) 
+\, 2\, \nabla_\m \le(J \, \pa^\m \f\ri)\Big] \,,
\eea
where
\be \label{PMT-dW}
W_\f \equiv\, \fr{dW}{d\f} =\, - \, 2 \L \, \fr{dY}{d\f} =\, \fr{32 \L} s \, 
Y^2 (\f) \,\,,
\ee
for the form of $Y(\f)$ given above in Eq.\,(\ref{PMT-YL}).

\section{Extended MMT Cosmological evolution in the standard setup} \label{sec:cosmPMT}

In the standard spatially flat FRW space-time background, the mimetic constraint 
$X = 1$ as usual implies that the mimetic field could be treated as a dimensionally 
rescaled {\em cosmic clock} (viz. $\f \equiv t$, the co-moving time)\footnote{Actually, 
$\f \equiv \k t$. The dimensional factor $\k = \sq{8 \pi G_N}$ is however set to unity 
in this paper.}, without loss of generality
\cite{CM-mm}.
This obviously means $\fp \equiv \tp$, the present age of the universe. Henceforth, we shall use 
$t$ only, to refer to the mimetic field, as well as the comoving time coordinate, and always denote 
the total derivative with respect to $t$ by an overhead dot (e.g. $\dot Y \equiv \rfraa{dY}{\!dt}$). 

\subsection{Extended MMT cosmological equations} \label{sec:PMTsetup}

Note first that the vanishing of the (pseudo-)tracefree torsion mode $\cQ^\a_{~\m\n}$ 
[{\it cf}. Eq.\,(\ref{Q-constr})] is perfectly consistent with the stringent constraints
on the torsion tensor required for the FRW metric structure in the $U_4$ space-time
\cite{ssasb-tcons}.  
In fact, such constraints also imply that only the temporal components of the other 
torsion modes $\cT_\m$ and $\cA_\m$ should exist, and hence by Eqs.\,(\ref{T-constr}) 
and (\ref{A-constr}), the scalar fields $\f$ and $\c$ should only depend on time. 
Secondly, because of the form of $J_{\m\n}$ [{\it cf}. Eq.\,(\ref{J-def})], that stems 
out of the $\a (\square \f)^2$ term in the action (\ref{PMT-ac2})), the expression 
(\ref{em-d}) for $\Td_{\m\n}$ cannot in general be recast in the form of a perfect 
fluid energy-momentum tensor. Nevertheless, in the FRW space-time one has the 
relationship
\be \label{Box-phi} 
\square \f \,=\, -\, 3 \, H \,\,,
\ee
where $\, H = \dfrac{\dot a} a$ is the Hubble parameter, with $a (t)$ denoting the scale factor. Therefore, as is well-known  
\cite{CMV-mm,MV-mm,SVM-mm},
the Friedmann and Raychaudhuri equations (that could be derived from Eqs.\,(\ref{PMT-eq1})
and (\ref{PMT-eq2})) reduce to their standard forms (viz. the ones in presence of a 
perfect fluid matter) only with the effective potential $W (t)$ rescaled by a constant 
factor that depends on the coefficient $\a$ of the $(\square \f)^2$ term in the mimetic
action. Given the form of our MMT potential (\ref{PMT-W}), this simply implies a rescaling 
of the constant parameter $\L$ (which we shall consider as a {\em cosmological constant}):
\be \label{W-rescale}
\L \, \rightarrow \le(1 +\, 3 c_s^2\ri) \L \,\,, \quad \mbox{with} \quad
c_s^2 =\, \fr \a {2 - 3 \a} \,\,.
\ee
The quantity $c_s$ here is nothing but the sound speed of the mimetic field perturbations,
which the $\, \a (\square \f)^2$ term in the mimetic action famously leads to
\cite{CMV-mm,MV-mm,SVM-mm}.

Consider now the cosmological matter to be the (pressureless) baryonic dust, specifically
for the study of the late-time evolution of the universe:
\be \label{em-mat}
\Tm\,\!^\m_{~\n} = \diag \!\le[-\rmt, 0, 0, 0\ri] \,, 
\ee 
with the corresponding energy density given by 
\be \label{matdens}
\rmt (t) =\, \fr{\rmp}{a^3 (t)}\,\,, \quad \mbox{where} \quad 
\rmp \equiv \rmt\Big\vert_{t = \tp} .
\ee
The total effective energy density and pressure, $\r$ and $p$, which satisfy the 
standard Friedmann and conservation equations, viz.
\be \label{cosm-eqs}
H^2 =\, \fr{\r} 3 \qquad \mbox{and} \qquad \dot\r =\, -\, 3 H \, (\r + p) \,\,,
\ee
are then, respectively,
\bea \label{PMT-denspr}
&& \r =\, \rmt + \rdt =\, R +\, 3 p \,\,, \\ 
&& p =\, \pdt =\, - \, \mfrac 1 2 \, W(t) =\, - \L \Big[1 - Y (t)\Big] \,.
\eea
These of course reduce to that for the effective $\L$CDM solution in Paper\,1, once 
the BI field is turned off, i.e. in the limit the parameter $s \rightarrow 0$, whence 
$Y (t) \rightarrow 0$. In general however, (for $s \neq 0$) it is reasonable to  
consider the MMT energy density $\rdt$ to consist of a pressureless (dust-like) cold 
dark matter (CDM) density $\rct$, and a left-over $\rx$, which may be treated as the 
effective dark energy (DE) density. In other words, while seeking a MMT cosmological 
dark sector, we may conveniently decompose its energy density as
\be \label{PMT-dens}
\rdt (t) =\, \rct (t) +\, \rx (t) \,\,, 
\ee
with
\be \label{CDMdens}
\rct (t) =\, \fr{\rcp}{a^3 (t)} \,\,, \quad \mbox{where} \quad 
\rcp \equiv \rct\Big\vert_{t = \tp} .
\ee
The effective MMT fluid pressure $\pdt$ (that equals the total pressure $p$), on 
the other hand, may inevitably be attributed to that of the effective DE constituent, 
i.e.
\be \label{DE-pr}
\px (t) =\, p (t) =\, \pdt (t) =\, - \L \le[1 - Y (t)\ri] \,.
\ee
Hence, from Eqs.\,(\ref{cosm-eqs}) it follows that 
\be \label{DE-consv}
{\dot\r}_{_X} (t) = \, -\, 3 H (t) \, \big[1 + \wx (t)\big] \rx (t) \,\,,
\ee
for a barotropic DE equation of state (EoS):
\be \label{DE-EoS}
\wx (t) :=\, \fr{\px (t)}{\rx (t)} =\, -\, \fr{\L \le[1 - Y (t)\ri]}{\rx (t)} \,\,.
\ee

\subsection{Cosmic Super-acceleration in the extended MMT scenario} \label{sec:PMTsupaccl}

Eqs.\,(\ref{DE-consv}) and (\ref{DE-EoS}) lead to the first order differential equation
\bea \label{DE-dEoS}
\dot{\sw}_{_X} (t) &:=& - \bigg[\fr{\dot Y (t)}{1 - Y (t)} \nn\\
&& \qquad -\, 3 \, \Big\{1 + \wx (t)\Big\}\, H (t)\bigg]\, \wx (t) \,.
\eea
Now, given the form 
\be \label{Yt} 
Y (t) =\, \fr s {16 \le(t + s\ri)} \,\,, 
\ee
we have 
\bea
&& \dot Y (t) =\, - \fr s {16 \le(t + s\ri)^2} = - \, \fr{16} s \, Y^2 (t) \,\,,
\label{DYt} \\
&& \fr 1 {1 - Y (t)} =\, \fr{16 \le(t + s\ri)}{16 t + 15 s} \,\,. \label{InvYt}
\eea
Hence, for the constant parameter $s$ to be presumably positive-valued\footnote{For 
$s < 0$, the identification $\f \equiv t \equiv s \c^2$ implies an {\it imaginary}
BI field. We prefer to keep that possibility out of the scope of this paper, although 
it is not unphysical, and in some sense reminiscent of Ashtekar's original works on 
the canonical quantum gravity formulation
\cite{asht1-QGR,asht2-QGR}.
}, $\dot Y$ is negative definite, whereas $(1 - Y)^{-1}$ is positive definite. Also, 
in an expanding universe, the Hubble parameter $H$ is positive definite. Therefore 
in Eq.\,(\ref{DE-dEoS}), the first term within the square brackets is negative 
definite, and so is the second term, as long as $\wx > -1$. Since, in any viable DE 
model, the corresponding EoS parameter $\wx$ is negative-valued, at least from 
reasonably high red-shifts ($\sim 100$ or so) all the way to the distant future, we 
may safely say that $\dot{\sw}_{_X} < 0$ whenever $\wx > -1$. The inferences that 
can be drawn from this are as follows:
\ben[(i)]
\item  If we have a {\em non-phantom} (i.e. accelerating, but not {\em super-accelerating})
regime in a moderately distant past, i.e. $\wx > -1$, then with the progress of time $\wx$ 
would decrease and tend to attain the value $-1$, as the slope $\dot{\sw}_{_X}$ is always 
negative. In fact, at $t = t_{_C} \,$, the epoch at which $\wx$ equals $-1$, the slope 
$\dot{\sw}_{_X} < 0$. Thereafter (i.e. with the advancement of time further beyond 
$t_{_C}$), $\dot{\sw}_{_X}$ would continue to remain negative, albeit with progressively 
lesser magnitude, at least for a certain period after which it may become positive. Hence, 
given the circumstances, we would definitely have a {\it crossing} of the {\em phantom 
barrier} (or the $\wx = -1$ point) at the epoch $t = t_{_C}$. There of course remain 
issues of sorting out finer details, for e.g. the condition for such a crossing to happen 
in the past (i.e. $t_{_C} < \tp$), the possibility of the phantom regime to be 
{\em transient} (i.e. $\dot{\sw}_{_X}$ becomes positive after a certain lapse of time 
beyond $t_{_C} \,$, and increases to the extent that $\wx \geq -1$ once more), and so on. 
However, addressing to these issues requires us to solve Eq.\,(\ref{DE-dEoS}) explicitly 
by setting up the appropriate boundary conditions, or(and) assert somehow the appropriate 
range of values of the parameter $s$. Of course, appropriateness here implies satisfying 
the requirement $t_{_C} > t_{_T}$, where $t_{_T}$ denotes the epoch of transition from a 
decelerating phase to an accelerating phase. Otherwise, our presumption of having a 
non-phantom regime in the past, and consequently the phantom barrier crossing, would be 
invalidated.
\item If, on the contrary, we have a phantom (or, super-accelerating) regime in a 
moderately distant past, i.e. $\wx < -1$, then $\dot{\sw}_{_X}$ would be $< 0 \,$ ($> 0$) 
depending on whether the first term within the square brackets in Eq.\,(\ref{DE-dEoS}) is 
bigger (smaller) than the second term therein. Evidently, there would be no phantom 
barrier crossing for $\dot{\sw}_{_X} < 0 \,$, whereas for $\dot{\sw}_{_X} > 0 \,$ such a 
crossing can happen, albeit from a phantom phase to a non-phantom phase. In either case 
though, no realistic cosmological evolution could be extracted unless we have a prior 
knowledge of a decelerating phase, followed by an accelerating but non-phantom phase, to 
precede the super-accelerating regime that we assume. This means that our consideration 
must have to be diverted back to the point (i) above, in order that a viable phantom 
crossing MMT model can emerge.
\een
The bottomline is therefore that the extended MMT setup we are considering here would
definitely give rise to a super-accelerating cosmological scenario, via a fractional 
reduction in the effective potential $W (t)$ by an amount $Y (t)$, over and above the 
cosmological constant $\L$ we have had in Paper\,1. Most importantly, this 
super-acceleration is not apparently stemming out of any ghost (or phantom) degree(s) 
of freedom in the effective action (\ref{PMT-ac}). It is the result of the specific 
inverse time-dependence $Y(t)$ acquires [{\it cf}. Eq.\,(\ref{Yt})], due to our 
chosen coupling $\b(\f) \sim \f^2$, of the mimetic field and the Holst term, as well 
as the identification (\ref{PMT-id}). In fact, as we shall explain below, there is 
nothing unusual with the energy conditions, at least as long as the strength of $Y(t)$, 
measured by the parameter $s$ in Eq.\,(\ref{Yt}), is reasonably weak. After all, it is 
this parameter $s$ which decides not only the extent of the super-acceleration, but 
also the physical realizability of the same (see the next section for further 
clarification). A large value of $s$ compared to the original MMT coupling strength 
$\betp$ ($= \rfraa{\L \tp} 3$) necessarily means a strong super-acceleration, 
culminating from a large distortion of the $\L$CDM solution, that we had in Paper\,1 
because of the constant potential $\, \bW = 2 \L$ therein. Since any major deviation 
from $\L$CDM is not even remotely supported by the observations, we may safely assume 
the parameter $s$ to be small, and for all practical purposes treat $Y(t)$ as a 
perturbation over $\, \bW = 2 \L$. This may also be reasoned from a purely theoretical 
standpoint, as follows: 

Recall that by Eq.\,(\ref{Yt}), the fractional correction $Y(t)$ is a monotonically 
decreasing function of time. Its maximum value is only $\mfrac 1 {16}$, and that too 
is attained at $t = 0$. Therefore, at least for the late time cosmology, we may expect 
$Y(t)$ not to make any strong impact on the $\L$CDM solution one gets in its absence. 
Such an expectation would possibly surmount to the level of a conviction once we 
consider the smallness of $s$. More specifically, a small $s$ would imply that the 
solution of Eq.\,(\ref{DE-consv}) may not possibly change its limiting value, viz. 
$\rx\big\vert_{s=0} =\, \L\,$, by an amount of the order of $\L$ itself. Hence, $\L$ 
being the dominant term in it, $\rx$ would be positive-valued, and so would be the 
total density $\, \r \,(= \rmt + \rct + \rx)$. The positivity of $\rx$ may nonetheless 
seem a bit surprising, since it implies that our effective DE component is not 
characteristically phantom-like, yet it leads to a phantom barrier crossing. However, 
remember that such a component is liable to show some weirdness, as it is merely a 
mathematical artifact, meant only for facilitating our understanding of the cosmological 
dynamics in analogy with the conventional scalar field induced DE models.

Now, at least for small $s$, the phantom barrier would not be breached to that extent 
which would imply that the quantity $\,|1 + \wx| \sim \cO(1)$, i.e. $|\px|$ exceeding 
$\rx$ by an amount of the order of $\rx$ itself. This can be inferred right away from 
Eqs.\,(\ref{DE-pr}) and (\ref{Yt}) which lead to
\be \label{DEdenspr}
\rx (t) +\, \px (t) =\, \d\rx(t) +\, \fr{\L \, s}{16 \le(t + s\ri)}  \,\,,
\ee
where $\, \d\rx (t) =\, \rx(t) -\, \L \,$ denotes the correction in the DE density
over its limiting value $\L$. The second term on the right hand side of 
Eq.\,(\ref{DEdenspr}) is positive definite (under our presumption $s > 0$ of course).
Therefore, $(\rx + \px)$ would become negative only when $\d\rx$ becomes negative 
and exceeds that term in magnitude. However, as argued above, $\d\rx$ cannot be of 
the order of $\L$. So there is no question of $|\rx + \px|$ reaching up to the order 
of $\rx$. By all means then, the total equation of state parameter of the system, 
$\sw = \rfraa p \rho$, would not get reduced to a value $< -1$, even in the phantom 
regime. The reason is obvious --- in order to make $\sw < -1$, i.e. $(\rho + p) < 0$, 
the correction $|\d\rx|$ would need to overcome not only the second term in 
Eq.\,(\ref{DEdenspr}), but also $\rmt$ and $\rct$, which is impracticable at least 
for small $s$. We would thus have $\, \rho + p > 0$, in addition to $\rho > 0\,$. 
Therefore, both the {\em null} and {\em weak} energy conditions would hold
\cite{HE-econd,wald,poisson}.
Moreover, $\sw$ not crossing $-1$ means that it would have a fractional value, i.e. 
$|p| < \rho \,$. So the {\em dominant} energy condition would hold as well. 
Nevertheless, the {\em strong} energy condition, viz. $\,\rho + 3p > 0$, or $\sw 
> - \mfrac 1 3\,$, would be violated as usual for an accelerated cosmic expansion.

\section{Viable MMT Cosmology with Phantom barrier crossing} \label{sec:PMTphantcross}

Let us, for convenience, treat the inverse squared BI field as an effective coupling 
function, viz.
\be \label{sig}
\s (\t) :=\, \c^{-2} (\t) \,\,, \qquad \mbox{where} \quad \t = \fr t \tp \,\,.
\ee
Since $\, \c^{-2} = s\, \f^{-1} \equiv s\, t^{-1}\,$, by Eq.\,(\ref{PMT-id}), we can
 write
\be \label{sig-def}
\s (\t) =\, \fr \sgp \t \,, \qquad \mbox{where} \quad 
\sgp \equiv \s\big\vert_{\t = 1} = \fr 
s \tp \,\,.
\ee
Accordingly, Eq.\,(\ref{DE-dEoS}) can be recast in the form
\bea \label{DE-dEoS1}
\fr{d\wx}{d\t} &:=& - \le[\fr 1 {1 - Y (\t)} \, \fr{dY}{d\t} \ri. \nn\\
&& \le. \qquad \quad -\, 3 \tp \Big(1 + \wx (\t)\Big) H (\t)\ri] \wx (\t) \,,
\eea
where
\be \label{Y}
Y (\t) =\, \fr {\s (\t)} {16 \le[1 + \s (\t)\ri]} =\, \fr \sgp {16 \le(\t + \sgp\ri)} \,\,.
\ee
Therefore,
\bea 
&& \fr{dY}{d\t} =\, - \, \fr{16} \sgp  \, Y^2 (\t) \,\,, \\
&& \fr 1 {1 - Y (\t)} =\, \fr{16 \le(\t + \sgp\ri)}{16 \t + 15 \sgp} \,\,,
\eea
which when substituted back in Eq.\,(\ref{DE-dEoS1}) gives
\bea \label{DE-dEoS2}
\fr{d\wx}{d\t} &:=& \bigg[\fr \sgp {\le(\t + \sgp\ri) \le(16 \t + 15 \sgp\ri)} \nn\\
&& \quad +\, 3 \, \tp \big\{1 + \wx (\t)\big\} \, H (\t)\bigg] \, \wx (\t) \,.
\eea

\subsection{Effective Dark Energy state parameter in a Linear Approximation \label{sec:PMTDE}}

In order to solve Eq.\,(\ref{DE-dEoS2}), for the effective dark energy EoS parameter 
$\wx$, one first requires to find the functional form of $H (\t)$ by solving the 
cosmological equations (\ref{cosm-eqs}). While getting an exact analytical solution 
is always a hard proposition, numerical methods may be applied, with a lot of 
intuition though, from the point of view of setting priors on the parameters $\sgp, 
\tp$ and so on. We may however take a shorter course, since for the reasons given 
earlier, it suffices us to look for small deviations from the $\L$CDM scenario. Let 
us therefore resort to the following expansions:
\bea 
\wx(\t) &=& \wx^{(0)} (\t) + \sgp \wx^{(1)} (\t) + \sgp^2 \wx^{(2)} (\t) + \cdots \,,
\label{wx-pert} \\
H (\t) &=& H^{(0)} (\t) + \sgp H^{(1)} (\t) + \sgp^2 H^{(2)} (\t) + \cdots \,, 
\label{H-pert}
\eea
by assuming the parameter $s$ to be small enough, so that $\sgp = \rfraa s \tp$ can be 
treated as a perturbation parameter. The unperturbed quantities correspond to those for 
$\L$CDM, viz.\footnote{Throughout this paper, we denote the $\L$CDM parameters and 
functions by placing an overbar, for e.g. the $\L$CDM total energy density denoted by 
$\brho$, the corresponding total pressure denoted by $\bp$, and so on.} 
\be \label{0-pert}
\wx^{(0)} =\, - 1 \qquad \mbox{and} \qquad H^{(0)} (\t) =\, \bH (\t) \,\,,
\ee
whereas the linear order perturbation equation, obtained from Eqs.\,(\ref{DE-dEoS2}) - (\ref{H-pert}),
is given by
\be \label{wx1-eq}
\fr{d\wx^{(1)}}{d\t} \,=\, - \le[\fr 1 {16 \t^2} \,+\, 3 \tp \wx^{(1)} (\t)\, \bH (\t)\ri] \,.
\ee
For simplicity (and also the adequacy justified later on), we shall confine ourselves 
to this order of perturbation in what follows.

Consider first the zeroth order scenario, i.e. the effective $\L$CDM evolution in the
MMT cosmological setup. The total pressure being $\bp = - \L$, the corresponding 
Friedmann and conservation equations lead to following equation for the total energy 
density
\be \label{LCDM-dens-eq}
\fr{d\brho}{d\t} =\, - \, \tp \sq{3 \brho (\t)} \,\Big[\brho (\t) - \L\Big] \,,
\ee
with solution
\be \label{LCDM-dens}
\brho (\t) =\, \L \, \coth^2 \z (\t) \,\,,
\ee
where
\be \label{zeta}
\z (\t) =\, \zetp \t \,\,, \qquad \zetp =\, \z\big\vert_{\t = 1} = \, \fr{\tp \sq{3 \L}} 2 \,\,.
\ee
So, by the $\L$CDM Friedmann equation, the corresponding Hubble parameter assumes the form
\be \label{LCDM-t0-H}
\bH (\t) =\, \sq{\fr{\brho (\t)} 3} =\, \fr{2 \zetp}{3 \tp} \, \coth \le(\zetp \t\ri) \,.
\ee
Using this in (\ref{zeta}) we have, more conveniently,
\be \label{zeta-def}
\z (\t) = \tanh^{-1} \!\sq{\bOL (\t)} \,, \quad \zetp = \tanh^{-1} \!\sq{\bOLp} \,,
\ee
where $\,\bOL = \rfraa \L \brho \,$ is the $\L$-density parameter (specific to the 
zeroth order case), with value $\, \bOLp$ at the present epoch $t = \tp$ (or $\t = 1$).

Resort now to the linear (or first) order approximation, in which case we can determine 
\be \label{wx-lin}
\wx (\t) \approx\, -\, 1 +\, \sgp \wx^{(1)} (\t) \,\,,
\ee
by plugging Eq.\,(\ref{LCDM-t0-H}) back in Eq.\,(\ref{wx1-eq}), and solving the
resulting equation. This gives
\bea \label{wx1}
\wx^{(1)} (\t) &=& \fr 1 {16} \bigg[\fr 1 \t - \le(1 - 16 \, \sw_{_{X0}}^{(1)}\ri) 
\fr{\sinh^2 \zetp}{\sinh^2 \le(\zetp \t\ri)} \nn\\
&& \qquad -\, \fr{\zetp \le\{\Shi \le(2 \zetp \t\ri) - \Shi \le(2 \zetp\ri)\ri\}}
{\sinh^2 \le(\zetp \t\ri)}\bigg] \,,
\eea
where $\, \sw_{_{X0}}^{(1)} \equiv  \wx^{(1)}\Big\vert_{\t=1}\,$, and 
\be \label{Shi}
\Shi (x) := \int^x \fr{\sinh x'}{x'} \, dx' = \sum_{k = 0}^\infty \fr{x^{2k + 1}}
{(2k + 1)^2 \, (2k)!} \,,
\ee
is the hyperbolic sine integral function.

In order to get a quantitative measure of the extent to which $\wx$ deviates from the 
$\L$CDM value $\, \wx^{(0)} = -1$ at different epochs of time, it is convenient for
us to figure our the circumstances in which the time-evolution of the percentage change
\bea \label{pc-err}
\m (\t) &:=& \fr{\wx (\t) - \wx^{(0)} (\t)}{\wx^{(0)} (\t)} \, \times 100 \nn\\
&=& - \, 100 \, \Big[1 + \wx (\t)\Big] \,,
\eea
would comply with the observations. From Eqs.\,(\ref{wx-lin}), (\ref{wx1}) and 
(\ref{pc-err}) we see that at the present epoch $\t = 1$ (or $t = \tp$),
\be \label{wx10}
\sgp \, \sw_{_{X0}}^{(1)} \,\approx \, - \, \fr{\mup}{100} \,, \qquad \mbox{where} 
\quad \mup =\, \mu (\tp) \,.
\ee
Observational estimations of the value of $\wx (\tp)$, considering say, the well-known 
Chevallier-Polarski-Linder (CPL) ansatz 
\cite{CP-cpl,lind-cpl},
imply that the percentage correction $\mu (\tp) \equiv \mup$ to be typically $\cO (1)$ 
or lesser, up to the $1\s$ error limits. For instance, the Planck 2018 results of the 
analysis of CMB TT,TE,EE+lowE combined with Lensing, Baryon Acoustic Oscillation (BAO) 
and type Ia Supernovae (SNe) data show 
\cite{Planck18}
\be 
\wx (\tp) =\, -\, 1.028 \pm\, 0.032  \,\,,
\ee
at $68 \%$ confidence level. This means $\mup \in (-0.04,6)$, which indicates not only 
a cosmological evolution very close to $\L$CDM, but also a tendency of the universe to 
be in a super-accelerating (or phantom) state at least at low red-shifts. Therefore, 
while seeking a viable model of a dynamical dark phase of the universe, one requires 
to ponder heavily on prioritizing the demand that the time-evolution of $\wx$ must 
turn out to be slow enough, so that the $\L$CDM solution is not distorted too much at 
least at the later epochs. So our linear approximation is duly justified. Moreover, 
it is desirable to have a provision for a phantom barrier crossing at an epoch in the 
recent past regime, in order that the expansion of the universe super-accelerates, 
albeit mildly, at the present epoch. After all, allowing for such a crossing provides 
flexibility in making statistical estimations of the model parameters (using the
observational data), as opposed to say, the quintessence scenarios in which one has 
to comply with the condition $\wx > -1$ beforehand. As we shall see in the next 
subsection, our approximated solution for $\wx$ can indeed lead to the perception of 
a viable phantom crossing scenario, however at the expense of restricting the domain 
of the parameters $\sgp$ and $\mup$ severely.

\subsection{Parametric Bounds from Observational Results  \label{sec:PMTbounds}}

For illustrative purposes, let us resort to the results of Planck 2018 TT,TE,EE+lowE+Lensing+BAO combined 
analysis for the base $\L$CDM model 
\cite{Planck18},
as in Paper\,1. In particular, we shall use the best fit value of the estimated 
$\L$-density parameter at the present epoch, viz. $\bOLp = 0.6889$, which implies 
\be \label{zeta-estm}
\zetp =\, \tanh^{-1} \sq{\bOLp} =\, 1.1881 \,\,.
\ee
Now, it is obvious that the linear approximation would not lead to a significant 
effect on the deceleration-to-acceleration transition epoch $t_{_T}$. So to determine 
the realistic parametric range of $\sgp$ or $\mup$ with reasonable precision, it 
would suffice us to approximate $t_{_T}$ to be that for $\L$CDM\footnote{In fact, 
the deviation of the actual $t_{_T}$ from its $\L$CDM value turns out to be about 
$0.056 \,\sgp$, as shown rigorously by using the Planck 2018 TT,TE,EE+lowE+Lensing+BAO 
results in a subsequent paper
\cite{MMT-SD}.
With a presumably small $\sgp$ value, this deviation is really quite insignificant.}, 
which can be obtained simply from the condition for cosmic acceleration, viz. $\bw 
(\t_{_T}) = - \mfrac 1 3$, where $\t_{_T} \equiv \rfraa{t_{_T}} \tp$ and
\be \label{LCDM-EoS}
\bw (\t) :=\, \fr{- \L}{\brho (\t)} =\, -\, \bOL (\t) = \, - \, \tanh^2 \le(\zetp \t\ri) 
\ee
is the $\L$CDM total EoS parameter. With $\, \zetp = 1.1881$, it follows that 
\be \label{LCDM-tT}
\t_{_T} \equiv\, \fr{t_{_T}}\tp =\, \fr 1 \zetp \, \tanh^{-1} \sq{\fr 1 3} =\, 0.554 \,\,.
\ee
Moreover, from the parametric relation (\ref{wx10}), we see that the solution (\ref{wx1}) 
can admit a viable phantom barrier crossing only when, for a fixed $\sgp$, the range of 
$\mup$ is restricted, or vice versa. Of course, by `viable' we mean that such a crossing 
has to take place at an epoch $t_{_C}$ in the near past (or low redshift) era of the 
evolving universe, when the latter's expansion is already accelerating, after the end 
of the decelerating phase at $t = t_{_T}$. In other words, $t_{_C}$ must lie within the 
temporal range $\le(t_{_T},\tp\ri)$. 

Again, since the phantom crossing implies
\be \label{phantcross} 
0 = 1 + \wx (\t_{_C}) \approx \wx^{(1)} (\t_{_C}) \,, \quad \mbox{where} \quad 
\t_{_C} \equiv \fr{t_{_C}} \tp \,,
\ee
we have from the above equations (\ref{wx1}) and (\ref{wx10}),
\bea \label{viable-parm}
\mup &\approx& \fr{25 \,\sgp} 4 \, \fr{\zetp}{\sinh^2 \zetp} 
\bigg[\fr{\sinh^2 (\zetp \t_{_C})}{\zetp \t_{_C}} -\, \fr{\sinh^2 \zetp}{\zetp} \nn\\
&& \qquad \qquad \qquad -\, \Shi (2 \zetp \t_{_C}) +\, \Shi (2 \zetp)\bigg] \,.
\eea 
This relationship is crucial for determining the parametric bounds, as we shall see below.
%
\begin{figure}[htp]
\begin{subfigure}{\linewidth} \centering
   \includegraphics[scale=0.6]{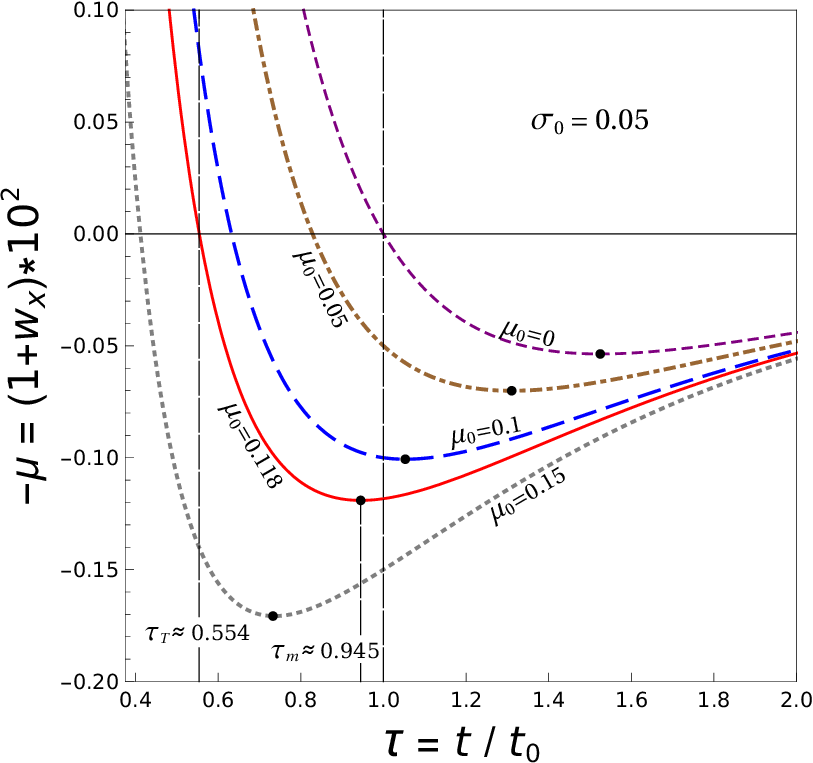}
   \caption{\footnotesize $-\m$ vs $t/\tp$ for $\sgp = 0.05$ and $\mup \in (0, 0.15)$.}
\end{subfigure}

\bigskip
\begin{subfigure}{\linewidth} \centering
   \includegraphics[scale=0.6]{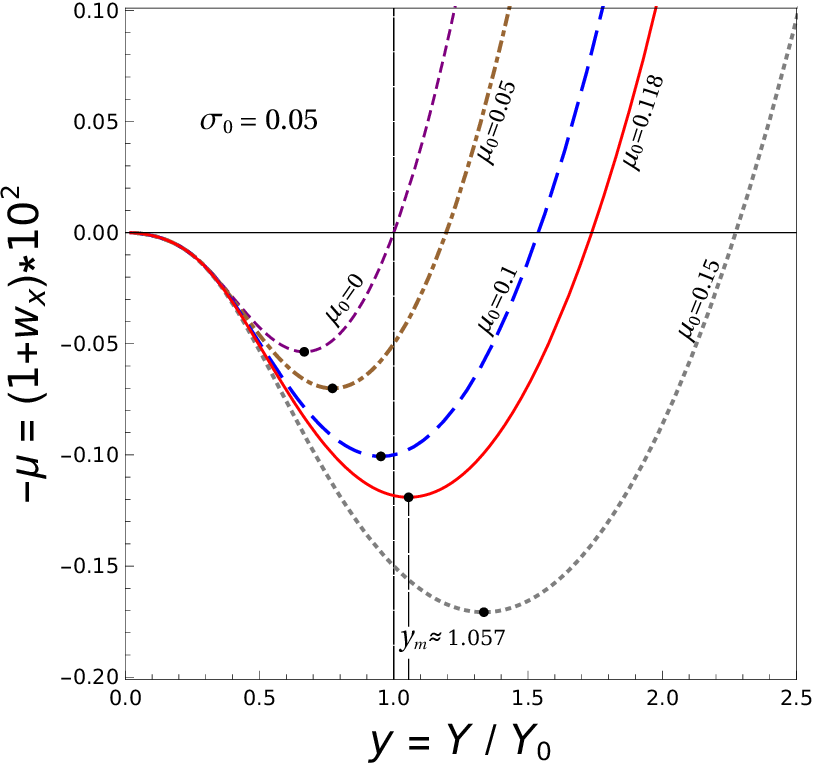}
   \caption{\footnotesize $-\m$ vs $Y/\Yp$ for $\sgp = 0.05$ and $\mup \in (0, 0.15)$.}
\end{subfigure}
\caption{\footnotesize Variation of the percentage change in the effective DE equation
of state: $- \mu = 100 (1 + \wx)$, with (a) the rationalized cosmic time $\t = t/\tp$, 
and (b) the rationalized correction $y = Y/\Yp$ in the MMT potential, for a fixed 
parametric value $\sgp = 0.05$ and discrete values of the parameter $\mup \in (0, 
0.15)$. The thick dots indicate the minima of the plots.}
\label{Mmt2-fig1}
\end{figure}
%

\vspace{5pt}
\no 
{\bf I.} For a fixed $\sgp$, the requisite parametric range of $\mup$ that meets the 
criterion $\, t_{_C} \in \le(t_{_T}, \tp\ri) \,$ is by no means broad, especially 
given the condition that $\sgp$ has to be small in order to make the linear 
approximation tenable. In particular, even if we consider fixing $\sgp$ at a 
reasonably significant value $\, 0.05$, the plots of the percentage correction 
$- \m (\t)$ [{\it cf}. Eq.\,(\ref{pc-err})] in Fig.\,\ref{Mmt2-fig1}\,(a), for 
various parametric values of $\mup$, make it evident that the latter cannot go much 
further beyond $0.1$. Otherwise (for e.g. when $\mup = 0.15$), $t_{_C}$ would fall 
below $t_{_T}$, thus making the phantom crossing unrealistic. Analytically, this can 
be easily seen from Eq.\,(\ref{viable-parm}), since with $\, \zetp = 1.1881 \,$ and 
$\, t_{_T} = 0.554 \, \tp \,$ [{\it cf}. Eqs.\,(\ref{zeta-estm}) and (\ref{LCDM-tT}) 
respectively], the criterion $\, t_{_C} \in \le(t_{_T}, \tp\ri) \,$ implies
\be \label{bound}
0 <\, \mup \,\lesssim\, 2.367 \, \sgp \,\le(=\, 0.118 \,,\,\, 
\mbox{for} \,\, \sgp = 0.05\ri) \,.
\ee
The solid curve in Fig.\,\ref{Mmt2-fig1}\,(a) shows the $\, - \m (\t)$ variation 
for the optimum value $\mup = 0.118$. The minimum (or turning) point of this 
curve, at $\, \t = \t_m = 0.945$, is of particular significance since it marks the 
earliest possible occurence of the minimum of $\, -\m (\t)$ corresponding to a 
$\mup$ value legitimate for a realistic phantom crossing. More specifically, at the 
crossing point $\t = \t_{_C}$ the function $\, - \m (\t)$ requires to be decreasing 
at such a rate that the minimum is reached very close to the present epoch, or 
afterwards. This is in fact a {\it general} requirement, irrespective of the chosen 
fixed value $\sgp = 0.05$, as verified below.

From Eqs.\,(\ref{wx-lin}) and (\ref{pc-err}) we see that the minimization of 
$\, - \m (\t)$ implies that of $\wx^{(1)} (\t)$. Therefore, using Eqs.\,(\ref{wx1-eq}),
(\ref{LCDM-t0-H}), (\ref{wx1}) and (\ref{wx10}), we get
\bea \label{tm-eq}
\fr{\tanh (\zetp \t_m)}{2 \, \zetp \t_m} &=& -\, 16 \, \t_m \, \wx^{(1)} (\t_m) \nn\\
&\approx& \fr{\zetp \t_m}{\sinh^2 (\zetp \t_m)} \bigg[\le(1 +\, \fr{4 \mup}{25 \sgp}\ri) 
\fr{\sinh^2 \zetp} \zetp \nn\\ 
&&\,\,+\, \Shi (2 \zetp \t_m) -\, \Shi (2 \zetp)\bigg] - \, 1 \,\,,
\eea
or, equivalently, by virtue of the relationship (\ref{viable-parm}),
\bea \label{tm-eq1}
\bigg[1 &+& \fr{\tanh (\zetp \t_m)}{2 \, \zetp \t_m}\bigg] \fr{\sinh^2 (\zetp \t_m)}
{\zetp \t_m} -\, \Shi (2 \zetp \t_m) \nn\\
&&\qquad \qquad \quad \approx\, \fr{\sinh^2 (\zetp \t_{_C})}{\zetp \t_{_C}} -\, 
\Shi (2 \zetp \t_{_C}) \,.
\eea
Solving this equation numerically, for $\zetp = 1.1881$, we find $\, \t_m = 0.945$ in 
the optimal case $\t_{_C} = \t_{_T} = 0.554$ (or equivalently $\mup = 2.367 \,\sgp \,$), 
without any allusion to the typical fixation of $\mup$ or $\sgp$.

It is also worth examining how the percentage correction $\, -\m$ varies with the 
quantity $\,y (t) \equiv \rfraa {Y (t)\!} \Yp \,$, where $\, \Yp \equiv Y(\tp)\,$. After 
all, it is the specific form of the fractional correction in the potential, $Y (t)$, 
given by Eq.\,(\ref{Y}), which makes the cosmic super-acceleration plausible. Now, as we 
see from Fig.\,\ref{Mmt2-fig1}\,(b), the criterion $\, t_{_C} \in \le(t_{_T}, \tp\ri)$, 
and correspondingly the viable range of $\mup$ (for a fixed $\sgp$), could be ascribed 
to the valid zone of occurence of $y = y_m \,$, i.e. the point at which $- \m$ is 
minimized. In particular, the above bound $\mup \lesssim 2.367 \,\sgp \,$ holds only 
for $\, y_m \lesssim 1.057 \,$, thus implying that the $-\m (y)$ variation should be 
such that the minimum point $y_m$ is reached no earlier than an epoch just surpassing
the present epoch. This is nonetheless a general criterion for the viability of the 
phantom crossing (not specific to the choice $\sgp = 0.05$). Fig.\,\ref{Mmt2-fig1}\,(b) 
also illustrates the {\em non-transience} of the phantom regime, i.e. the 
super-acceleration is ever-lasting (although it slows down progressively after 
reaching $y_m$). Analytically, this follows from Eq.\,(\ref{Y}) which implies $y \equiv
\rfraa {Y\!} \Yp \approx \t^{-1}$. Therefore, $y$ tends to vanish in the asymptotic 
limit ($\t \rightarrow \infty$), and so does the correction $\m$.  

\begin{figure}[htp]
\begin{subfigure}{\linewidth} \centering
   \includegraphics[scale=0.6]{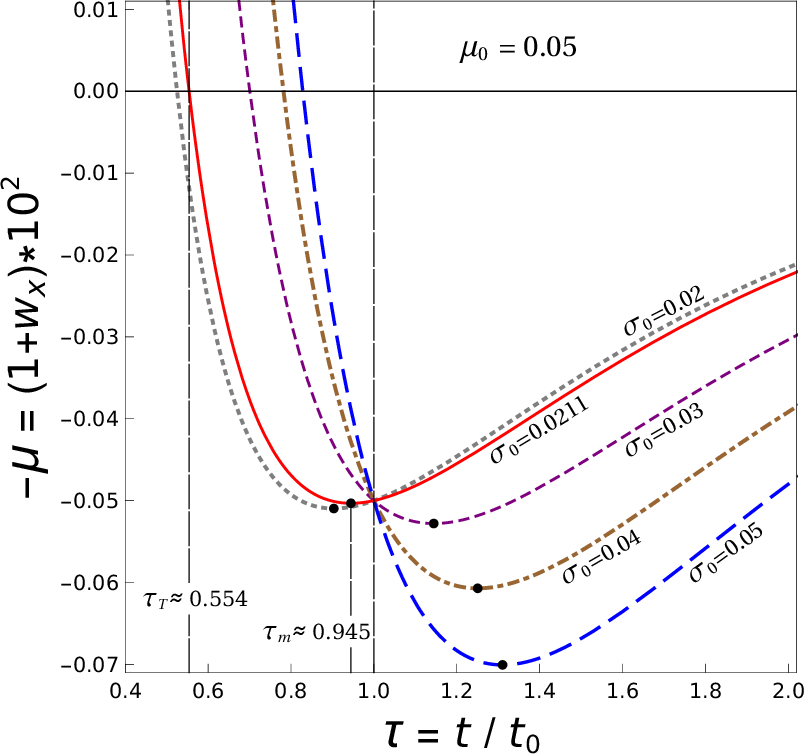}
   \caption{\footnotesize $-\m (t)$ vs $t/\tp$ for $\mup = 0.05$ and $\sgp \in (0.02, 
   0.05)$.}
\end{subfigure}

\bigskip
\begin{subfigure}{\linewidth} \centering
   \includegraphics[scale=0.6]{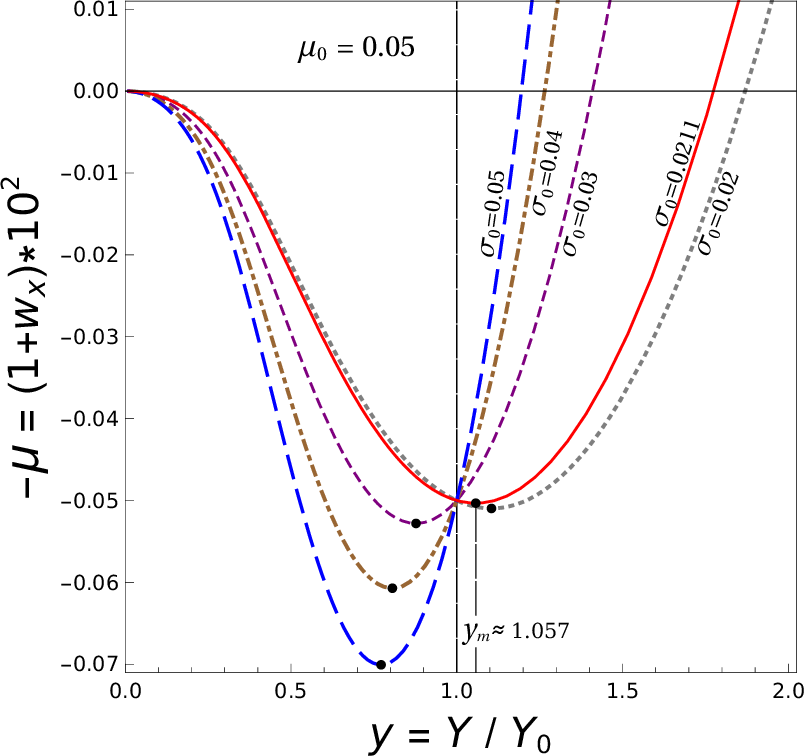}
   \caption{\footnotesize $-\m (t)$ vs $Y/\Yp$ for $\mup = 0.05$ and $\sgp \in (0.02, 
   0.05)$.}
\end{subfigure}
\caption{\footnotesize Variation of the percentage change in the effective DE equation
of state: $- \mu = 100 (1 + \wx)$, with (a) the rationalized cosmic time $\t = t/\tp$, 
and (b) the rationalized correction $y = Y/\Yp$ in the MMT potential, for a fixed 
parametric value $\mup = 0.05$ and discrete values of the parameter $\sgp \in (0.02, 
0.05)$. The thick dots indicate the minima of the plots.}
\label{Mmt2-fig2}
\end{figure}

\vspace{5pt}
\no 
{\bf II.} If, on the other hand, we resort to a fixed value of $\mup$, say a fairly low 
one ($= 0.05$), then as illustrated in Fig. \ref{Mmt2-fig2}\,(a), and as derived from 
Eq.\,(\ref{viable-parm}), a realistic phantom barrier crossing requires a lower limit on 
the parameter $\sgp$, close to (but greater than) $0.02$, for the $\zetp$ and $\t_{_T}$ 
estimated above. 
%
%
However, if we aspire to get a much larger $\mup$, say $\sim 0.5$ (i.e. about $\mfrac 
1 2 \%$ correction in $\wx$), then by Eq.\,(\ref{viable-parm}), we require $\sgp \gtrsim
0.211$, which may nonetheless compel us to go beyond the linear approximation. The above 
arguments concerning the criterion $\, t_{_C} \in \le(t_{_T}, \tp\ri) \,$ being ascribed
to the domain of appearance of the turning points $\t = \t_m$ would apply here as well. 
Also for completeness, it is worth examining the variation of $- \m$ with $y = \rfraa 
{Y\!} \Yp$, for $\mup$ fixed at $0.05$ and various parametric values of $\sgp$. The 
corresponding plots shown in Fig.\,\ref{Mmt2-fig2}\,(b) display the same non-transient 
phantom regime as illustrated in Fig. \ref{Mmt2-fig1}\,(b) above.


\section{Evolving Torsion parameters and the extent of the Super-acceleration}
\label{sec:PMT-TA}

Let us recall the expressions (\ref{T-evol}) and (\ref{A-evol}) for the relevant 
torsion parameters, viz. the norms of the torsion mode vectors $\cT_\m$ and $\cA_\m$. 
With the MMT coupling function $\b (\f)$ given by Eq.\,(\ref{PMT-coup}), and the 
identification $\, s \c^2 = \f \equiv t$, such expressions reduce to
\bea 
\cT^2 &\equiv& -\, g^{\m\n} \, \cT_\m \cT_\n =\, \fr 9 {t^2} \le[1 - 
\fr s {4 (t + s))}\ri]^2 \,, \label{T-norm}\\
\cA^2 &\equiv& -\, g^{\m\n} \, \cA_\m \cA_\n =\, \fr{9 s}{t (t + s)^2} \,\,.
\label{A-norm}
\eea
Our interest, of course, is in getting a quantitative measure of the extent to which 
the scenario in Paper\,1 is modified when the parameter $s \neq 0$. For this we 
require a careful examination of the time-evolution of    
\be \label{UTUA}
U_T :=\, \fr{\cT^2 - \bcT^2}{\bcT^2} \qquad \mbox{and} \qquad
U_A :=\, \fr{\cA^2}{\bcT^2} \,\,,
\ee
which are the fractional changes over the {\it only} existent torsion parameter in 
Paper\,1, viz. 
\be \label{LCDM-T}
\bcT^2 \equiv \Big[- g^{\m\n} \cT_\m \cT_\n\Big]_{s = 0} =\, \fr 9 {t^2} \,\,,
\ee
that had led to the $\L$CDM solution therein 
\cite{RAS-MMT}.

Using Eqs.\,(\ref{T-norm}), (\ref{A-norm}) and (\ref{LCDM-T}), along with the 
definitions (\ref{sig-def}), we can write
\bea
U_T (\t) &=& -\, \fr{U_A (\t)} 2 \le(1 + \fr{7 \sgp}{8 \t}\ri) \,, \label{UTt} \\
U_A (\t) &=& \fr{\sgp \t}{(\t + \sgp)^2} \,\,,  \label{UAt}
\eea
and consequently re-express the fractional correction to the effective potential 
$\bW = 2\L$ of Paper\,1 as
\be \label{YU}
Y (\t) =\, \fr \sgp {16 \le(\t + \sgp\ri)} =\, -\, \fr{U_A (\t) 
+\, 16\, U_T (\t)}{112} \,\,.
\ee
Note that so far there is no approximation --- all the relations 
(\ref{UTt})\,--\,(\ref{YU}) are exact. Nevertheless, from the perspective of 
our analysis in the previous section (anticipating per se, mild deviations 
from $\L$CDM due to the MMT extension), we shall only consider small values 
of the parameter $\sgp$ from here on.  
%
\begin{figure}[htp]
   \begin{subfigure}{\linewidth} \centering
     \includegraphics[width=3in,height=3in]{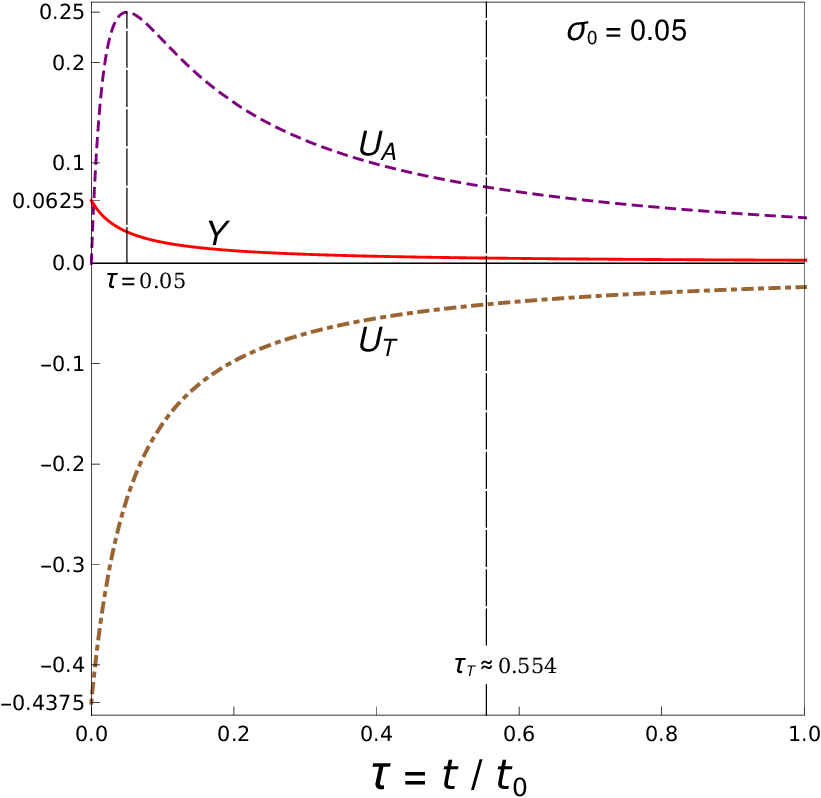}
     \caption{\footnotesize $U_A$, $U_T$ and $Y$ vs $t/\tp$ for $\sgp = 0.05$.}
   \end{subfigure}

\bigskip
   \begin{subfigure}{\linewidth} \centering
     \includegraphics[width=3in,height=3in]{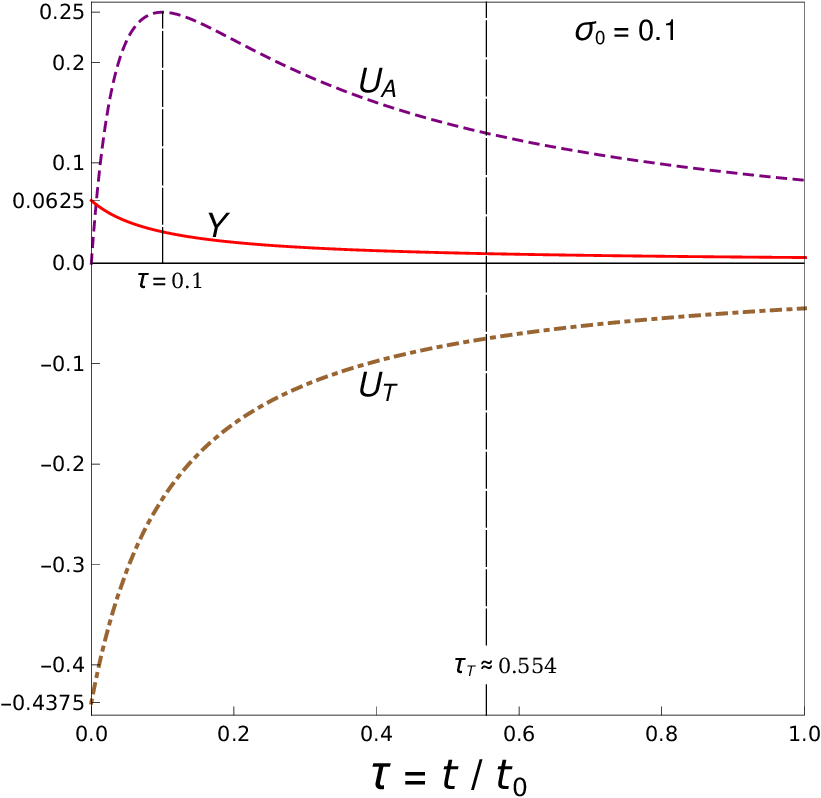}
     \caption{{\footnotesize $U_A$, $U_T$ and $Y$ vs $t/\tp$ for $\sgp = 0.1$.}}
   \end{subfigure}
\caption{\footnotesize Variations of the fractional changes in the torsion parameters, 
viz. $U_A (\t)$ and $U_T (\t)$, and their resultant $Y (\t)$, for fixed parametric 
values: (a) $\sgp = 0.05$ and (b) $\sgp = 0.1$.}
\label{Mmt2-fig3}
\end{figure}
%

Now for small $\sgp$, the fractional corrections $U_A$ and $U_T$, although decrease 
with time, remain of the same order of magnitude as $\sgp$ itself, in the temporal 
range of interest $\, t \in (t_{_T}, \tp) \,$, or $\, \t \in (\t_{_T},1)$. However, 
the resulting fractional change in the potential, $Y$, turns out be comparatively 
much smaller, as illustrated with two fiducial settings $\sgp = 0.05$ and $\sgp = 
0.1$ in Figs.\,\ref{Mmt2-fig3}\,(a) and (b) respectively. 
%
\begin{figure}[htp]
   \begin{subfigure}{\linewidth} \centering
     \includegraphics[width=3in,height=3in]{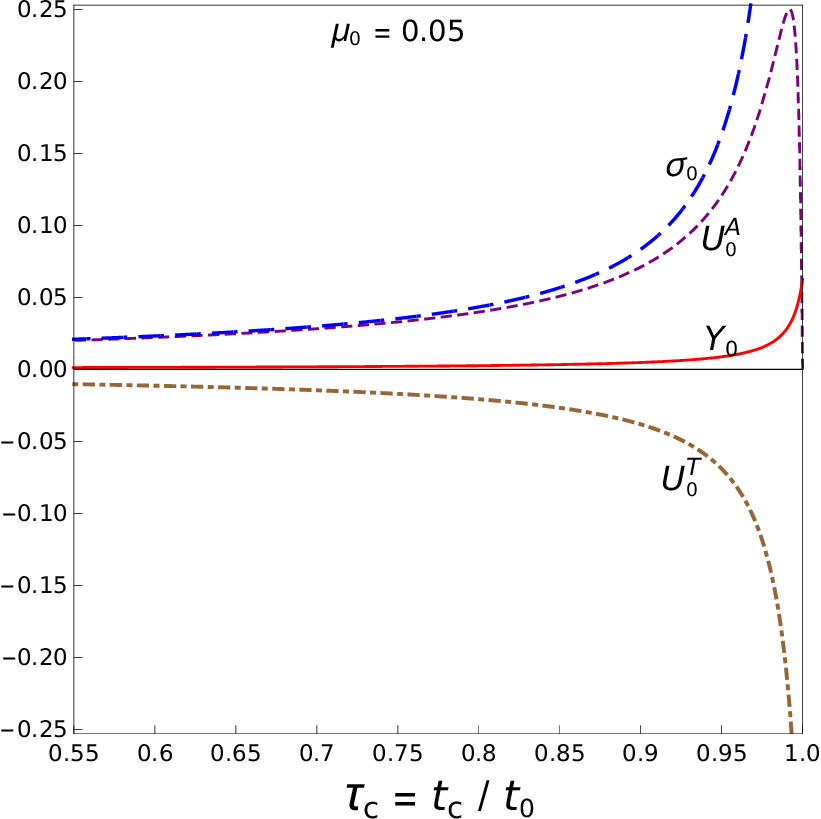}
     \caption{\footnotesize $\sgp$, $\UAp$, $\UTp$ and $\Yp$ vs $t_{_C}/\tp$ for 
     $\mup = 0.05$.}
   \end{subfigure}

\bigskip
   \begin{subfigure}{\linewidth} \centering
     \includegraphics[width=3in,height=3in]{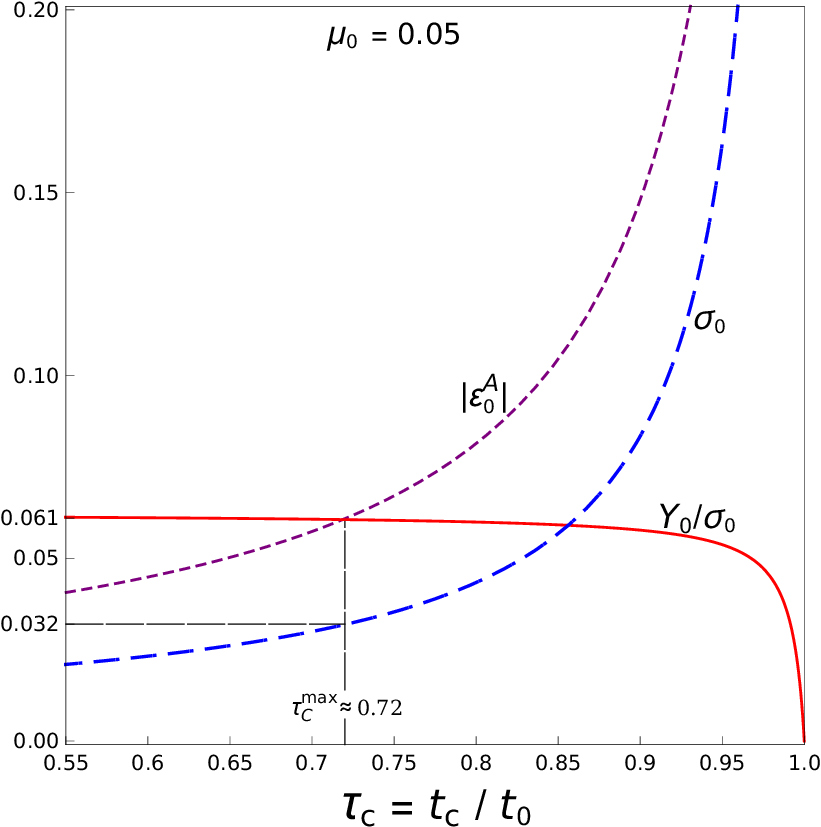}
     \caption{{\footnotesize $\sgp$, $\Yp/\sgp$ and $|\eAp|$ vs $t_{_C}/\tp$ for 
     $\mup = 0.05$.}}
   \end{subfigure}
\caption{\footnotesize Approximated variations of the parametric values of (a) $\sgp$ 
and the fractional changes in the torsion parameters at the present epoch, viz. $\UAp$ 
and $\UTp$, as well as their resultant $\Yp$, and (b) $\sgp$, $\Yp/\sgp$ and $\vert
\eAp\vert$, i.e. the fractional deviation of $\UAp$ from $\sgp$, with the rationalized 
phantom crossing time $\t_{_C} = t_{_C}/\tp$ in the viable range $(0.55,1)$, for a 
fixed $\mup = 0.05$.}
\label{Mmt2-fig4}
\end{figure}
%
This smallness of $Y$, and as a consequence the weak time-variation of the effective
DE equation of state parameter $\wx$, of course justifies the linear approximation in 
the preceding section. Note also that the peak values of $U_A, |U_T|$ and $Y$ are 
independent of the value of $\sgp$. In particular, the above expression for $U_A$ 
implies that the latter attains its maximum value $\,(= 0.25)$ at $\, \t = \sgp$, as 
illustrated in Figs.\,\ref{Mmt2-fig3}\,(a) and (b). On the other hand, $|U_T|$ and 
$Y$ being monotonically decreasing functions of $\t$, have their maximum values 
(equal to $0.4375$ and $0.0625$ respectively) at $\t = 0$.   

As to the values of $U_A$, $|U_T|$ and $Y$ at the present epoch $t = \tp$, it is 
evident from Eqs.\,(\ref{UTt})\,--\,(\ref{YU}) that all of them are smaller than 
$\sgp$, for $\, 0 < \sgp \lesssim 1\,$: 
\bea 
&& \UAp \equiv\, U_A (\tp) =\, \sgp \Big[1 +\, \eAp\Big] \,, 
\label{UAbound}\\ 
&& \le|\UTp\ri| \equiv\, |U_T (\tp)| =\, \fr \sgp 2 \Big[1 +\, \eTp\Big] \,,
\label{UTbound}\\ 
&& \Yp \equiv\, Y (\tp) =\, \fr \sgp {16} \Big[1 +\, \eYp\Big] \,, 
\label{Ybound}
\eea
where
\bea 
&& \eAp =\, -\, \fr{\sgp \le(2 + \sgp\ri)}{(1 + \sgp)^2} =\, - \, 2 \sgp + 
\cO (\sgp^2) \,\,, \label{UAdev}\\ 
&& \eTp =\, -\, \fr{\sgp \le(9 + 8 \sgp\ri)}{8 (1 + \sgp)^2} =\, - \, \fr{9 \sgp} 8 
+ \cO (\sgp^2) \,\,, \label{UTdev}\\ 
&& \eYp =\, -\, \fr \sgp {1 + \sgp} =\, - \, \sgp + \cO (\sgp^2) \,\,. \label{Ydev}
\eea
Now, using Eq.\,(\ref{viable-parm}) one can determine how $\sgp$, and hence 
$\, \UAp, \, |\UTp|$ and $\Yp$, vary approximately with the rationalized phantom 
crossing time $\t_{_C} \,(= \rfraa{t_{_C}} \tp)$, for a fixed value of the percentage 
correction $\mup$ in $\wx$ at $t = \tp$ (or $\t = 1$). Fig.\,\ref{Mmt2-fig4}\,(a) 
shows such parametric variations, within the viable range $t_{_C} \in (t_{_T},\tp)$, 
or $\t_{_C} \in (0.55,1)$, for a fiducial setting $\mup = 0.05$ (which nonetheless 
is well within the $1 \s$ error estimates from recent observations
\cite{Planck18}).

A stringent upper limit can in principle be placed on $t_{_C}$, and hence on $\sgp$, for 
a given value of $\mup$, from the argument that $\Yp$ would be of significance only when 
it is $\gtrsim \sgp |\eAp|$, where $|\eAp|$ is the absolute fractional deviation of 
$\UAp$ from $\sgp$. Specifically, this could be seen as a consequence of a stringent 
limit being imposed on the extent of the linear approximation, for which $\UAp \approx 
\sgp$, by Eq.\,(\ref{UAbound}). The higher order change, i.e. the deviation of the exact 
$\UAp$ from $\sgp$, given by the amount $\sgp |\eAp|$, need to be $\, \lesssim \Yp$, 
because otherwise $\Yp$ would also be a higher order effect, which would in turn imply 
the invalidity of the linear approximation\footnote{One can alternatively consider 
imposing $\Yp \gtrsim \sgp |\eTp|\,$ or $\, \Yp \gtrsim \sgp |\eYp|$, however the 
resulting upper bound on $\sgp$ (or $t_{_C}$) would then be less stringent, since both 
$|\eTp|$ and $|\eYp|$ are smaller than $|\eAp|$, by Eqs.\,(\ref{UAdev})\,--\,(\ref{Ydev}).}. 
As shown in Fig.\,\ref{Mmt2-fig4}\,(b), the effect of $\Yp$ of a very small amount 
($\lesssim 0.061 \sgp$), could yet be perceptible in the linear approximation (i.e. 
satisfy the condition $\Yp \gtrsim \sgp |\eAp|\,$) and lead to a reasonably significant 
amount of $\mup \, (= 0.05)$, only if $t_{_C} \lesssim 0.72 \tp$, or $\sgp \lesssim 
0.032$ correspondingly. In principle, a lesser upper bound on $\sgp$, culminating to an 
even smaller amount of $\Yp$, may provide a more significant $\mup$, however at the 
expense of reducing the upper limit of $t_{_C}$. One can therefore infer the following:
\bit
\item As the phantom crossing time $t_{_C}$ cannot be smaller than the 
deceleration-to-acceleration transition time $t_{_T}$, the percentage correction 
$\mup$ (in the effective DE equation of state $\wx (\tp)$) cannot get enhanced 
much further beyond the value $0.05$, at least in the linear approximation.
\item Nevertheless, the extended MMT-cosmological scenario seems more attractive 
for a $t_{_C}$ value closer to $t_{_T}$, i.e. for an earlier commencement of the 
super-accelerating regime, from the point of view of having larger value of $\mup$ 
with lesser strength of coupling $\sgp$ of the mimetic field with the Holst term. 
In other words, the weaker the MMT extension, the stronger is the effect in 
$\wx (\tp)$. There is a limit to this of course, since $t_{_C}$ must not fall 
short of $t_{_T}$.
\eit

\section{Conclusion} \la{sec:concl}

A viable phantom crossing evolution of a unified cosmological dark sector is thus
demonstrated by extending the basic mimetic-metric-torsion (MMT) formalism with an 
explicit coupling of the mimetic field $\f$ and the Holst term, motivated from the 
following:
\bed
\item Firstly, its compliance with the basic precept of MMT gravity, viz. the
preservation of conformal symmetry while letting $\f$ to manifest geometrically 
as the source of torsion (or certain mode(s) thereof).
\item Secondly, and most crucially, its instrumentality in making torsion's main 
characteristic, viz. the  anti-symmetry, accountable for the dynamical evolution of a 
given physical system.
\eed
To be more specific, the reason why we have incorporated the $\f$-coupled Holst term 
is that, apart from keeping the conformal symmetry of the MMT theory unaffected, it 
ideally meets our objective of perceiving a plausible effect of the axial (or 
pseudo-trace) mode $\cA_\m$ of torsion, in addition to that of the latter's trace 
mode $\cT_\m$, on the cosmic evolution profile. In particular, for clarity in 
interpreting results, irrespective of the explicit $\f$-coupling, we have preferred 
to resort to a non-topological characterization of the {\it bare} Holst term by 
promoting its coefficient, viz. the Barbero-Immirzi (BI) parameter $\c$, to the 
status of a field. Considering this field to be a (presumably {\it primordial}) 
pseudo-scalar (or, an axion), we could avoid gravitational parity violation, and 
hence a direct confrontation with the latter's miniscule observational signature at 
cosmological scales. Consequently, an identification of $\f$ with $\c^2$ (i.e. the 
simplest possible even function of $\c$), using a scalar Lagrange multiplier field, 
had made it evident that both the (a\,priori independent) torsion modes $\cT_\m$ 
and $\cA_\m$ are induced by $\f$. In the paradigm of the standard FRW cosmology, 
the outcome is in some sense, a `geometric unification' of the dark sector, as both 
the effective DM and DE components of the universe, being purportedly the artefacts 
of $\cT_\m$ and $\cA_\m$, have had their dynamics driven by the same scalar field 
$\f$.

Technically of course, the MMT theory (introduced in Paper\,1) is designed to have 
almost all of its weight resting on the pre-assigned contact coupling, $\, \b (\f)$, 
between the mimetic field and the entire torsion-dependent part of the Riemann-Cartan 
($U_4$) Lagrangian. In the extended formulation, such a 
coupling obviously carries to the Holst term, which augments the $U_4$ Lagrangian. 
The further postulation of a pseudo-scalar BI field, and subsequently the
identification $\f = s \c^2$ (where $s$ is a dimensionless constant), are only 
expected to lead to a subtle modification of what $\, \b (\f)$ alone can inflict 
on a given system configuration. Nevertheless, there arises the important question 
as to whether such a modification, despite its suppressed quantitativeness, can 
turn out to be significant in the cosmological context. This is what we have 
addressed specifically via our analysis in the preceding sections.

In particular, taking the MMT coupling function in the form $\, \b (\f) \sim \f^2\,$,
we have brought the time-variance of the effective mimetic potential $W (\f)$ in the
reckoning, as opposed to its constancy in Paper\,1, viz. $W = \bW = 2 \L$, due to the 
only existent torsion mode $\cT_\m$ (in absence of the Holst term) therein. While 
the coupling $\, \b (\f) \sim \f^2\,$ has had its motivation from the point of view 
of its natural appearance in metric-torsion theories involving scalar field(s)
\cite{shap-trev,ssasb-mst1,ssasb-mst2,ssasb-mst3},
the effect of the typical functional form it ascribes to the fractional modification
$Y(\f)$ of the constant potential $\bW = 2 \L$ (of Paper\,1) has turned out to be 
quite fascinating in the cosmological context. Specifically, such a $\f^2$-coupling 
implies an inverse functional dependence of $Y$ on $\f$, or equivalently, on the 
cosmic time $t$ in the FRW framework. The consequence of this, by virtue of the 
mimetic constraint, is a {\it super-accelerating} phase of cosmic evolution, marked 
by the diminution of the effective DE equation of state parameter $\wx (t)$ below 
the $\L$CDM value $-1$. In fact, as we have inferred via a set of arguments, the 
super-acceleration is always plausible in our extended MMT scenario, which most 
importantly, does not appear to have any theoretical obstacles, such as those 
concerning ghost or phantom degrees of freedom. This may be understood from the 
point of view that none of the energy conditions, except the strong one, is violated 
in our entire formalism. The violation of the strong energy condition is of course 
nothing unusual --- it is mandated for any attempt of explaining the late-time 
cosmic acceleration (regardless of an even later super-acceleration) in the standard 
FRW framework.

Nevertheless, despite having the theoretical consistency, there remained the task 
for us to assert whether the transition from $\wx > -1$ to $\wx < -1$, or the {\it 
phantom barrier crossing}, is at all physically realizable, and if so, under what 
condition(s). In specific terms, a physically realistic (or viable) phantom crossing 
implies that the epoch of its occurence, $t = t_{_C}$, has to be within the time-span 
$(t_{_T},\tp)$, where $t_{_T}$ denotes the deceleration-to-acceleration transition 
epoch and $\tp$ denotes the present epoch. While the lower bound $t_{_C} > t_{_T}$ 
is obvious, the upper bound $t_{_C} < \tp$ actually pertains to a fair amount of 
observational signification of a {\it mild} super-acceleration to have commenced in 
the near past phase of evolution of the universe, and continuing to the present epoch 
and beyond
\cite{Planck18}.
The mildness of course alludes to the gross observational concordance on $\L$CDM 
cosmology. That is to say, in whichever way the DE equation of state parameter $\wx$ 
may deviate from the $\L$CDM value $-1$, observations disfavour the latter's error 
limits to be breached to any perceivable extent.  Now, to get the condition(s) for 
$t_{_C} \in (t_{_T},\tp)$ to hold (if at all), it is imperative to determine the 
functional form of $\wx (t)$ by explicitly solving the corresponding cosmological 
equations. However, instead of attempting an exact analytic solution, we have 
adopted an approximation method, upon treating the effect of the function $Y(\f)$ 
as a small perturbation over that due to the constant potential $\bW = 2 \L$, viz. 
the $\L$CDM solution in Paper\,1. Admittedly, the strong observational support for 
$\L$CDM has made it reasonable for us to assume that $Y(\f)$ would always inflict 
a small deviation of $\wx$ from the value $-1$.

Working out $\wx (t)$ in the approximated form, we have shown that the phantom 
crossing is indeed physically realizable, however for a severely restricted range 
of values of our MMT model parameters, viz. $\sgp$ and $\mup$, where $\, \sgp = 
\rfraa s \tp$ and $\mup$ is the fractional amount by which $\wx$ differs from $-1$ 
at $t = \tp$. In particular, limiting our analysis to the {\em linear} order of 
the (presumably small) parameter $\sgp$, we have illustrated the time-evolution of 
$\wx$ for certain fiducial settings of either a fixed $\sgp$ or a fixed $\mup$. 
Strikingly enough, all such settings have made the revelation of the phantom regime
being {\it non-transient}, i.e. the super-acceleration is ever-lasting, even though 
it slows down progressively after reaching an optimum point. A close inspection of 
the evolution profiles of the torsion parameters, viz. the norms of the torsion 
trace and pseudo-trace mode vectors $\cT_\m$ and $\cA_\m$, have enabled us to 
determine the validity of the linear approximation, and hence the extremity of the 
phantom crossing time $t_{_C}$, as well as that of $\wx(\tp)$ (or equivalently, 
$\mup$). In fact, the latter is found to be not exceeding a value typically well 
within the Planck 2018 TT,TE,EE+lowE+Lensing+BAO error estimates for $\L$CDM
\cite{Planck18}.
This nonetheless signifies a very low extent of the super-acceleration, in well 
accord with our presumption of its smallness.

On the whole, we have seen some really appealing outcomes of our entire program of
extending the basic MMT formalism of Paper\,1. Not only that it consolidates the 
tantalizing picture of a geometrically unified dark sector we have had therein, but 
also discerns the viability of the cosmic super-acceleration, or the phantom phase, 
without any impending danger of dealing with a ghost-like entity. This may 
nonetheless be counted towards the robustness of the emergent cosmological model, 
since after all, from a purely theoretical standpoint, a provision for the phantom 
barrier crossing is desirable on account of the flexibility thus comprehended while 
making statistical estimations of the model parameters using the observational data. 

Nevertheless, apart from the technical difficulties the extreme mildness of the 
cosmic super-acceleration may pose in its detection with the current generation of 
observational probes, there remain several issues to be pondered on. The foremost,
of course, are those concerning the Ostrogradsky ghost or(and) gradient instabilities 
at the perturbative level, due to the presence of higher derivative ($\sim \le(\square
\f\ri)^2$) extension of the MMT Lagrangian
\cite{IRS-minst,HNK-minst,ZSML-minst}.
A plausible removal of such instabilities is demonstrated, via a proposed alternative 
extension, in the Appendix below. Secondly, there is the associated problem of the 
caustic singularities, which in fact affects not only the mimetic theory, but also 
any scalar-tensor model exhibiting the mimicry of a dust-like fluid description. 
Caustic formation is unavoidable for any constrained scalar field, such as the mimetic 
one, that leads to a dust velocity flow tangential to the time-like geodesics
\cite{barv-mm,CR-mm,ramz-mm,BR-mm}.
Within the scope of mimetic gravity, this problem has been circumvented via vectorial 
(gauge field) extensions 
\cite{GMF-mgf,GMFM-mgf,JV-mgf}.
However, in the MMT scenario the presence of the torsion coupling with the mimetic
field $\f$ may be of some aid, since it implies that apart from being a fundamental
mode of gravity, the field $\f$ manifests itself geometrically as the source of one 
(or more) vector mode(s) of torsion. Furthermore, the geodesics of actual importance
here are the affine geodesics (or, the {\em auto-parallel} curves) which are in
general not identical to the metric geodesics in the Riemann-Cartan ($U_4$) space-time.
Therefore, a careful inspection resulting in a chosen specific form of the 
$\f$-coupling with the torsion vector mode $\cT_\m$ had enabled us to get an 
effective dust-like (MMT) fluid description in Paper\,1 (see the Appendix therein).
Admittedly however, a more rigorous contemplation of the focusing theorem and caustics
is needed in this context, by taking note of the existent references 
\cite{LM-CS,PO-CS},
which is one of the assignments we have undertaken presently.

Other pertinent questions are, for instance,
(i) can there be any obvious way of realizing the $\f$-coupling with torsion, with 
or without involving the Holst term?
(ii) would that be perceptible via the Hilbert-Palatini formulation of mimetic gravity? 
(iii) how would the magnitude of the super-acceleration be affected if there be a 
constraint more complicated than (\ref{PMT-id})? 
(iv) can that constraint be imposed automatically (i.e. without the aid of a Lagrange 
multiplier) via some mechanism? 
Attempts of addressing to some of these are currently underway, and hopefully be reported 
soon.

\section*{Acknowledgement}
The authors acknowledge useful discussions with Mohit Sharma. The work of AD is supported by University Grants Commission (UGC), Government of India.

\section*{Appendix: A note on the possible aversion of ghost and gradient instabilities} 
\label{Appdx-MMT-ext}

Let us refer back to the action (\ref{PMT-ac}). Apart from the matter action $\Sm$, it has the pure MMT action with the general structure
\be \label{PMT-ac-str}
\Smmt =\, \int \! d^4 x \sq{- g} \le[\cL^{(g)} +\, \cL^{(c)} +\, \cL^{(h)}\ri] \,,
\ee 
that includes the purely gravitational part (alongwith the $\f$-coupled torsional
contributions) described by the Lagrangian
\be \label{PMT-grav}
\cL^{(g)} =\, \fr 1 2 \le[R +\, \b (\f) \le(\Th +\, \fr{~^\star \Th}{2\c}\ri)\ri] \,, 
\ee
the constraint part with Lagrangian 
\be \label{PMT-constr}
\cL^{(c)} =\, \mfrac 1 2 \Big[\l \cdot (X - 1) +\, \n \cdot \!\le(\f - s \c^2\ri) \Big] \,,
\ee 
and the higher derivative (HD) augmentation, described by the Lagrangian
\be \label{PMT-hd}
\cL^{(h)} =\, \mfrac \a 2 \le(\square \f\ri)^2 \,.
\ee 
As is well-known, such an augmentation crucially leads to a non-zero sound speed of 
the mimetic matter density perturbations, given by
\cite{CMV-mm}
\be \label{snd}
c_s =\, \sq{\fr \a {2 - 3 \a}} \,\,,
\ee
however, at the expense of making the mimetic field dynamical. While a non-zero
$c_s$ is essential for defining the usual quantum fluctuations due to $\f$, so that 
the latter can provide the seeds of the observed large scale structure of the universe
\cite{CMV-mm,SVM-mm},
the presence of a HD term in the action is in general problematic as it leads to
Ostrogradsky instabilities. Remarkably, the mimetic constraint $X = 1$ comes to the 
rescue here, as it eliminates the dynamical mode arising from $\le(\square \f\ri)^2$, 
and in turn the Ostrogradsky ghost, for a suitable range of values of the parameter 
$\a$
\cite{CMV-mm}.
However, a rather critical inspection reveals that there remains a stringent problem 
of a gradient instability, which requires to be confronted with substantial rigour
\cite{IRS-minst,HNK-minst,ZSML-minst}.
As demonstrated by several authors in recent times, such a problem could be remedied 
by introducing an explicit contact coupling of an appropriate function of $\square \f$ 
(and possible other HD constructs, e.g. $\f_{\m\n} \f^{\m\n}\,$ where $\, \f_{\m\n} 
\equiv \nabla_\m \nabla_\n \f$) with curvature invariants, such as 
$R$
\cite{HNK-minst,ZSML-minst,TK-minst,FGM-minst}.
Nevertheless, this still implies a dynamical $\f$ --- the assumption which we 
endeavor to disregard in what follows.

To be specific, we consider a scenario in which the pure MMT action (\ref{PMT-ac-str})
has the same gravitational and constraint Lagrangians given respectively by 
(\ref{PMT-grav}) and (\ref{PMT-constr}), but the HD Lagrangian (\ref{PMT-hd}) 
replaced by
\be \label{PMT-hd-alt}
\cL^{(h)} =\, \mfrac 1 2 \Big[\f \,\square \psi \,-\, m^2 \psi^2 +\, n \psi R \,-\, 
\ell^2 R^2\Big] \,,
\ee 
where $\psi$ is a dynamical scalar field (considered dimensionless), and $n, \ell, m$
are constants ($n$ is dimensionless, whereas $\ell$ and $m$ have unit length and mass
dimensions, respectively). Eq.\,(\ref{PMT-hd-alt}) essentially means that we are
considering a theory with two scalar fields --- one is the non-dynamical mimetic
field $\f$ whereas the other is a (presumably primordial) dynamical field with a
non-canonical kinetic term and having explicit contact couplings with both $\f$ and
$R$. The purpose of incorporating the $R^2$ term is to eliminate any scale dependence
in the theory, as we shall see below. Note also that although the proposed Lagrangian 
(\ref{PMT-hd-alt}) seems to have nothing to do with the presence of torsion, a 
correspondence could be traced, albeit with certain additional subtleties which are 
being examined in our subsequent works (in preparation)
\cite{MMT-DS,MMT-Stab}.
For the time being, however, in this paper, we consider the Lagrangian 
(\ref{PMT-hd-alt}) to be just an exemplary one, that replaces Eq.\,(\ref{PMT-hd}) 
in an attempt of providing a non-zero $c_s$, and side-by-side removing the 
gradient instability, without requiring the mimetic field to be dynamical, to 
begin with.

Eliminating a surface term one can make $\psi$ an auxiliary field, whence the 
corresponding variation of the full action would lead to the constraint
\be \label{psi}
\psi =\, \fr{\square \f +\, n R}{2 m^2} \,\,.
\ee 
Hence, implementing this and the other constraint (\ref{PMT-id}), and proceeding
in the same way as in section \ref{sec:genPMT}, we get
\bea \label{PMT-ac2a}
\Smmt &=& \fr 1 2 \int \! d^4 x \sq{- g} \bigg[(1 \,+\, \e\, \square \f)\, R 
\,+\, \a \le(\square \f\ri)^2 \nn\\
&&\qquad \qquad \qquad +\, \l \, (X - 1) -\, W(\f) \, X \bigg] \,,
\eea
where $W (\f)$ is as given by Eq.\,(\ref{PMT-W}), and we denote 
\be \label{PMT-coeffs}
\a \equiv\, \fr 1 {4 m^2} \quad \mbox{and} \quad \e \equiv\, \fr \ell {2 m} \,\,,
\ee 
upon choosing $\, n = 2 m \ell\,$, so that the $R^2$ contribution gets cancelled 
(thereby removing the scale-dependence).

Let us now resort to the standard Arnowitt-Deser-Misner (ADM) decomposition of the 
line element for the perturbed FRW space-time
\be \label{ADM}
ds^2 = - N^2 dt^2 +\, h_{ij}\! \le(dx^i + N^i dt\ri)\! \le(dx^j + N^j dt\ri) ,
\ee 
where $h_{ij}$ is the induced metric, and $N$ and $N^i$ are respectively the lapse
function and the shift vector. It is convenient to choose the unitary gauge in
which the scalar (mimetic) field perturbations $\d \f =\, 0 \,$, and the induced 
metric assumes the form
\be \label{unit-gauge}
h_{ij} (\vx,t) =\, a^2 (t) \,\exp \Big[2 I_{ij} \z (\vx,t) +\, \c_{ij} (\vx,t)\Big] \,,
\ee 
where $\z$ and $\c_{ij}$ respectively denote the scalar (or curvature) mode and
the elements of the (transverse and trace-free) tensor mode of perturbations. As
$\f$ is purely a function of time in this gauge, the mimetic constraint $X = 1$ 
implies
\be \label{MC}
N =\, \dot \f \,\,, \quad \pa_i N =\, 0 \,\,.
\ee
Setting therefore, without loss of generality, $N = 1$, as in ref.\cite{ZSML-minst}, 
we get
\be \label{K}
\square \f =\, -\, K \,\,, 
\ee
where $\, K =\, h^{ij} K_{ij} \,$ denotes the trace of the extrinsic curvature
$\,K_{ij} = \mfrac 1 2 \le[{\dot h}_{ij} - \le(\nab_i N_j + \nab_j N_i\ri)\ri]$.

The perturbed MMT action can hence be written, after certain simplifications, as
\bea \label{PMT-ac-pert}
\Smmt = \fr 1 2 \!\int\! d^4 x \sq{h} \bigg[\big(1 &-& \e K\big) \Big\{\!\Rin 
+ K_{ij}K^{ij}\Big\} \nn\\
&-&\! \le(1 - \a\ri) K^2 -\, W (\f)\bigg] ,
\eea
where $h$ is the determinant of the induced metric $h_{ij}\,$ and $\Rin$ denotes 
the corresponding curvature scalar. Varying this action with respect to the shift
vector $N^i$ we get the corresponding equation of motion, given up to the first
order in the perturbations, as
\be \label{Shift-EoM}
\pa_i N^i =\, \fr{2 \e\, \pa^2 \z + \le[2 \le(1 - \e K\ri) - 3 \le(\a - 5 \e H\ri)
\ri] a^2}{\le(\a - 5 \e H\ri) a^2} \,,
\ee 
which shows that $N^i$ only has a longitudinal part
\cite{ZSML-minst}. 
Following then the steps of ref.\cite{ZSML-minst}, and the general methodology of 
ref.\cite{mald-Pert}, we obtain the quadratic order action (after a great deal of 
simplifications) as\footnote{The detailed calculations could be found in the 
subsequent papers
\cite{MMT-DS,MMT-Stab},
which we hope to report soon.} 
\bea \label{PMT-ac-pert2}
S^{(2)}_{_{MMT}} = \int\! d^4 x \,\, a^3 \bigg[\! &f^{(T)}&\! \le\{\dot{\c}_{ij}
\dot{\c}^{ij} - e^{(T)} \pa_i \c_{jk} \pa^i \c^{jk}\ri\} \nn\\
&+&\!\! f^{(S)} \le\{{\dot \z}^2 - e^{(S)} \, \pa_i \z \pa^i \z\ri\} \nn\\
&+&\!\! \tilde{f}^{(S)} \! \le(\pa^2 \z\ri)^2 -\, \mfrac 1 2 \, W(\f)\bigg] \,,
\eea 
where 
\bea 
f^{(T)} &=& \fr{1 - 3 \e H} 8 \,, \label{f-coeff-T}\\
e^{(T)} &=& a^{-2} \,, \label{e-coeff-T}\\
f^{(S)} &=& \fr{\le(1 - 3 \e H\ri) \Big[3 \le(\a - 3 \e H\ri) - 2\Big]}
{\a - 5 \e H} \,, \label{f-coeff-S}\\
e^{(S)} &=& - \Big[3 \le(\a - 3 \e H\ri) - 2\Big]^{-1} \bigg[\le(\a - 3 \e H\ri) \nn\\
&& \,-\, 2 \e^2 \le(\fr 3 {1 - 3 \e H} -\, \fr 5 {\a - 5 \e H}\ri) \dot H\bigg] 
a^{-2} \,, \label{e-coeff-S}\\
\tilde{f}^{(S)} &=& -\, \fr{2 \e^2 a^{-4}}{\a - 5 \e H} \,, \label{f-coeff-S1}
\eea 
are the purely time-dependent coefficients of the tensor (T) and scalar (S) mode 
kinetic terms.

Ghost-freeness of the theory requires $f^{(T)}$, $f^{(S)}$ and $\tilde{f}^{(S)}$ to be 
positive definite at all epochs, and the gradient instabilities would be removed if
$e^{(T)}$ and $e^{(S)}$ remain positive definite as well. Moreover, from Eq.\,(\ref{snd})
we see that $\, 0 \leq c_s^2 \leq 1 \,$ implies $\, 0 \leq \a \leq \mfrac 1 2$. Therefore,
severe restrictions are bound to get imposed on the Hubble parameter $H (t)$, after 
being compounded with the demand that $\dot H (t) < 0 ~(\forall \,\, t)$, so that by the 
Raychaudhuri-Friedmann equation $\, \r + p > 0 \,$, i.e. the null and weak energy 
conditions hold (see the discussion in the last paragraph of section \ref{sec:cosmPMT}).
A close inspection of Eqs.\,(\ref{f-coeff-T})\,--\,(\ref{f-coeff-S1}) nonetheless reveal 
that all the above criteria are fulfilled for the following bounds on the dimensionless 
Hubble rate
\be \label{phys-bounds}
\fr \a {5 \e \Hp} \,\leq\, \fr H \Hp \,\leq\, \fr \a {3 \e \Hp} \,\,.
\ee 
So, to have the range of values of $H/\Hp$ enhanced for a given $\a \, \in \le[0, 
\mfrac 1 2\ri]$ and $\Hp$, so as to bring more flexibility in the model, we require
the parameter $\e \ll \a\,,$ in units of $\Hp$. The smallness of $\e$ implies a very 
weak coupling of $\square \f$ with $R$, for which of course the background MMT
cosmological evolution would not be affected much. Hence, our analysis in sections
\ref{sec:cosmPMT}, \ref{sec:PMTphantcross} and \ref{sec:PMT-TA} would still be valid,
by and large.

\end{document}